\pdfoutput=1
\documentclass{article}

\PassOptionsToPackage{square,comma,compress}{natbib}
\usepackage[main, final]{template/neurips_2026}

\usepackage{amsmath,amsfonts,bm}

\def\1{\bm{1}}

\def\va{{\bm{a}}}
\def\vb{{\bm{b}}}

\def\vf{{\bm{f}}}

\def\vh{{\bm{h}}}

\def\vo{{\bm{o}}}

\def\vq{{\bm{q}}}

\def\vs{{\bm{s}}}

\def\vw{{\bm{w}}}
\def\vx{{\bm{x}}}
\def\vy{{\bm{y}}}
\def\vz{{\bm{z}}}

\def\vdelta{{\bm{\delta}}}

\def\vtheta{{\bm{\theta}}}

\def\vtau{{\bm{\tau}}}

\def\vomega{{\bm{\omega}}}

\def\mD{{\bm{D}}}

\def\mI{{\bm{I}}}

\def\mW{{\bm{W}}}

\DeclareMathAlphabet{\mathsfit}{\encodingdefault}{\sfdefault}{m}{sl}
\SetMathAlphabet{\mathsfit}{bold}{\encodingdefault}{\sfdefault}{bx}{n}

\newcommand{\R}{\mathbb{R}}

\usepackage{float}                          
\usepackage[section,above,below]{placeins}  
\renewcommand{\floatpagefraction}{0.99}     

\usepackage{caption}                        
\DeclareCaptionLabelSeparator{bar}{ $|$ }
\usepackage{hyperref}                       
\usepackage{url}                            
\usepackage{chngcntr}                       
\usepackage{xcolor}                         
\definecolor{blue}{HTML}{1B7BC7}
\definecolor{red}{HTML}{CF3C33}
\definecolor{yellow}{HTML}{E7A341}
\definecolor{green}{HTML}{06A66C}
\definecolor{purple}{HTML}{9856B8}
\definecolor{orange}{HTML}{E27136}
\definecolor{cyan}{HTML}{23C1BB}
\definecolor{magenta}{HTML}{E163B1}
\definecolor{lime}{HTML}{82BE54}
\definecolor{indigo}{HTML}{8488E1}
\hypersetup{
  colorlinks=true,
  citecolor=blue,
  linkcolor=blue,
  urlcolor=blue,
}
\newcommand{\figref}[2][]{\hyperref[#2]{\figureautorefname~\ref*{#2}#1}}
\newcommand{\panel}[1]{(\textbf{#1})}
\usepackage{etoolbox}
\makeatletter
\patchcmd{\hyper@makecurrent}{%
    \ifx\Hy@param\Hy@chapterstring
        \let\Hy@param\Hy@chapapp
    \fi
}{%
    \iftoggle{inappendix}{
        \@checkappendixparam{chapter}%
        \@checkappendixparam{section}%
        \@checkappendixparam{subsection}%
        \@checkappendixparam{subsubsection}%
        \@checkappendixparam{paragraph}%
        \@checkappendixparam{subparagraph}%
    }{}%
}{}{\errmessage{failed to patch}}

\newcommand*{\@checkappendixparam}[1]{%
    \def\@checkappendixparamtmp{#1}%
    \ifx\Hy@param\@checkappendixparamtmp
        \let\Hy@param\Hy@appendixstring
    \fi
}
\makeatletter

\newtoggle{inappendix}
\togglefalse{inappendix}

\apptocmd{\appendix}{\toggletrue{inappendix}}{}{\errmessage{failed to patch}}

\usepackage{etoolbox}

\makeatletter

\pretocmd{\NAT@citex}{%
  \let\NAT@hyper@\NAT@hyper@citex
  \def\NAT@postnote{#2}%
  \setcounter{NAT@total@cites}{0}%
  \setcounter{NAT@count@cites}{0}%
  \forcsvlist{\stepcounter{NAT@total@cites}\@gobble}{#3}}{}{}
\newcounter{NAT@total@cites}
\newcounter{NAT@count@cites}
\def\NAT@postnote{}

\def\NAT@hyper@citex#1{%
  \stepcounter{NAT@count@cites}%
  \hyper@natlinkstart{\@citeb\@extra@b@citeb}#1%
  \ifnumequal{\value{NAT@count@cites}}{\value{NAT@total@cites}}
    {\ifNAT@swa\else\if*\NAT@postnote*\else%
     \NAT@cmt\NAT@postnote\global\def\NAT@postnote{}\fi\fi}{}%
  \ifNAT@swa\else\if\relax\NAT@date\relax
  \else\NAT@@close\global\let\NAT@nm\@empty\fi\fi
  \hyper@natlinkend}
\renewcommand\hyper@natlinkbreak[2]{#1}

\patchcmd{\NAT@citex}
  {\ifNAT@swa\else\if*#2*\else\NAT@cmt#2\fi
   \if\relax\NAT@date\relax\else\NAT@@close\fi\fi}{}{}{}
\patchcmd{\NAT@citex}
  {\if\relax\NAT@date\relax\NAT@def@citea\else\NAT@def@citea@close\fi}
  {\if\relax\NAT@date\relax\NAT@def@citea\else\NAT@def@citea@space\fi}{}{}

\makeatother

\usepackage[utf8]{inputenc}
\usepackage[T1]{fontenc}
\usepackage{hyperref}
\usepackage{url}
\usepackage{booktabs}
\usepackage{amsfonts}
\usepackage{nicefrac}
\usepackage{microtype}
\usepackage{graphicx}
\usepackage{wrapfig}
\usepackage{makecell}
\usepackage{enumitem}

\newcommand{\website}{\url{https://rat-ncap.github.io/}}
\newcommand{\webref}[1]{\href{https://rat-ncap.github.io/}{#1}}

\title{Neural Circuit Architectural Priors\\for Rat Locomotion}

\author{%
  Nikhil~X.~Bhattasali\textsuperscript{1}\thanks{Equal contribution.} ,\enspace
  Jaron~Cui\textsuperscript{1}\footnotemark[1] ,\enspace
  Lerrel~Pinto\textsuperscript{1},\enspace
  Grace~W.~Lindsay\textsuperscript{2,3}\\
  \textsuperscript{1}Department of Computer Science,\enspace
  \textsuperscript{2}Department of Psychology,\enspace
  \textsuperscript{3}Center for Data Science\\
  New York University\\
  \texttt{nikhilxb@nyu.edu}\\
}

\begin{document}

\maketitle

\begin{abstract}
Animals generate behavior through continuous interactions between their neural circuits, bodies, and environment.
Embodied simulations provide a means to study these interactions by testing neural circuit models in biologically realistic settings.
Recent efforts have developed embodied simulations of diverse animals, and emerging work on biologically grounded neural architectures has begun to integrate circuit connectivity and single-unit dynamics from experimental data into artificial neural networks, introducing stronger inductive biases and closer mechanistic alignment with biology.
How can we develop embodied simulations that enable the study of increasingly complex bodies and neural circuits?
In this work, we address the challenge through coordinated advances in the simulated body, neural architecture, and training infrastructure.
We introduce \textit{Rat v2}, a biomechanical model that provides an efficient and flexible tool for simulating rat behavior.
To control this body, we develop \textit{Rat NCAP}, a biologically grounded neural architecture for locomotion based on mammalian spinal circuits that incorporates cell-type-specific connectivity and a novel flexor-extensor neuromechanical interface.
To optimize circuit parameters, we develop GPU-accelerated simulation and training infrastructure.
The integrated system learns to locomote at different speeds and transition between gaits without imitation learning, producing locomotion statistics, kinematics, and cell-type-specific neural activity consistent with experimental data.
Systematic analyses reveal that the neuromechanical interface substantially affects trainability and gait quality, and \textit{in silico} cell-type ablations reproduce known gait phenotypes and predict gait changes under graded perturbations.
Together, these findings support the view that embodied simulations at an intermediate level of abstraction can simultaneously reproduce complex behavior and yield mechanistic insights into the neural circuits that generate it.

\end{abstract}

\section{Introduction}
\label{sec:1_introduction}

Animals generate behavior through continuous interactions between their neural circuits, bodies, and environment.
Embodied simulations provide a means to study these interactions by testing neural circuit models in biologically realistic settings, thereby complementing and extending experimental work~\citep{Ausborn2021ComputationalModelingSpinal}.
Recent efforts have developed simulated body models of diverse animals, including the nematode~\citep{Sarma2018OpenWormOverviewRecent}, fly~\citep{Lobato-Rios2022NeuroMechFlyNeuromechanicalModel,Wang-Chen2024NeuroMechFlyV2Simulating}, salamander~\citep{Ijspeert2007SwimmingWalkingSalamander}, mouse~\citep{Ramalingasetty2021WholebodyMusculoskeletalModel}, and rat~\citep{Merel2019DeepNeuroethologyVirtual}.
These models span a wide range of complexity and biorealism, raising a fundamental question about abstraction:
How much biological detail should be incorporated into body models and neural circuit models if we aim to simultaneously reproduce complex behavior and gain mechanistic insights?\linebreak
At the detailed end, we can build neuromusculoskeletal bodies with overactuated Hill-type muscular control, driven by conductance-based neurons arranged in biologically grounded circuits~\citep{Ramalingasetty2024AllFours3D}.
This approach provides mechanistic insights and captures fine-grained phenomena that are crucial for certain questions, but the resulting models can be costly to simulate, difficult to optimize, and often underconstrained by available data, especially for complex behaviors.
At the abstract end, we can use simplified bodies with joint position control, driven by standard artificial neural networks~(ANNs) trained with reinforcement or imitation learning~\citep{Merel2019DeepNeuroethologyVirtual,Aldarondo2024VirtualRodentPredicts}. 
This approach is more tractable to optimize and can scale to complex behaviors, but trained model mechanisms are difficult to relate to experimentally identified neural circuits, and producing naturalistic behavior often relies on imitation learning, which is biologically unrealistic and requires costly data collection.
This motivates embodied simulations at an intermediate level of abstraction that balance scientific fidelity with computational tractability.

ANNs offer desirable features as neural models by abstracting away complex biophysical details like spikes and ion channels, and they can be flexibly trained to reproduce behavior and neural activity.
Emerging work has begun to integrate biological structure from experimental data into ANNs, including circuit connectivity and single-unit dynamics.
In the fly, where connectomic data is especially complete~\citep{Dorkenwald2024NeuronalWiringDiagram, Scheffer2020ConnectomeAnalysisAdult}, this approach has yielded models that can reproduce and predict neural activity with cell-type specificity, while maintaining the trainability and abstraction of ANNs~\citep{Lappalainen2024ConnectomeconstrainedNetworksPredict,Duan2025SynapsesDynamicsObtaining}.
In mammals, this approach is challenging because relevant neuronal and circuit information must be integrated from diverse sources, but a long history of anatomical, physiological, and genetic studies has characterized key circuits with sufficient resolution to motivate structured neural models.

In previous work on Neural~Circuit~Architectural~Priors~(NCAP), \cite{Bhattasali2024NeuralCircuitArchitectural} showed that mammalian spinal cord circuits can constrain a biologically grounded neural architecture with cell-type-specific connectivity to control quadruped locomotion.
The architecture introduced stronger inductive biases than standard ANNs, leading to efficient training and robust gaits in addition to closer mechanistic alignment with biology.
However, that study used a simplified quadruped body with 12 actuated degrees of freedom, designed to isolate architectural principles rather than match biological reality.
Scaling to biorealistic bodies matters for neuroscience, as richer bodies enable direct comparison to animal data and more convincing validation of circuit hypotheses.
Yet achieving this scaling is far from straightforward.
Biorealistic bodies are inherently harder to control, with higher-dimensional action spaces, more complex dynamics, and narrower stability margins.
They also raise new questions about the neuromechanical interface---the way in which neural activity maps to body actuation, which can include joint torques, joint positions, or muscle activations.
Finally, these bodies are costly to simulate and train, slowing hypothesis testing.
How can we develop embodied simulations that enable the study of increasingly complex bodies and neural circuits?

In this work, we address the challenge through coordinated advances in the simulated body, neural architecture, and training infrastructure.
We introduce \textit{Rat~v2}, a biomechanical rat body with 38 actuated degrees of freedom based on~\cite{Merel2019DeepNeuroethologyVirtual} that improves anatomical fidelity while serving as a more efficient and flexible tool through GPU-accelerated simulation and a modular environment builder. 
To control this body, we develop \textit{Rat~NCAP}, a biologically grounded neural architecture for locomotion that scales mammalian spinal architectural priors from~\cite{Bhattasali2024NeuralCircuitArchitectural} to the rat by incorporating more complete neural connectivity and a novel flexor-extensor neuromechanical interface that enables more muscle-like control.
To optimize circuit parameters, we develop GPU-accelerated simulation and training infrastructure that makes evolutionary optimization in this body efficient.
Together, these components enable the study of increasingly complex bodies and neural circuits within embodied simulations at an intermediate level of abstraction.

The trained \textit{Rat~NCAP} produces locomotion at various speeds within a single model and transitions naturalistically between gaits.
Importantly, the model produces naturalistic behavior without imitation learning, demonstrating the power of architectural priors to guide training.
The trained model produces locomotion statistics, kinematics, and cell-type-specific neural activity consistent with experimental data.
Systematic analyses reveal that the neuromechanical interface substantially affects trainability and gait quality.
Finally, \textit{in silico} cell-type ablations reproduce known gait phenotypes and predict gait changes under graded perturbations.
Together, these findings support the view that embodied simulations at an intermediate level of abstraction can simultaneously reproduce complex behavior and yield mechanistic insights into the neural circuits that generate it.

\begin{samepage}
The key contributions of this work are:

\begin{enumerate}[leftmargin=0.25in, nosep, parsep=1ex, itemsep=0ex]
    \item \textbf{Develop the biologically grounded \textit{Rat~NCAP} controller from locomotion circuits.}
    \item \textbf{Validate trained behavior and neural activity against experimental data on rat locomotion.}
    \item \textbf{Evaluate how neuromechanical interface affects trainability and gait quality.}
    \item \textbf{Probe cell-type contributions and generate testable predictions via \textit{in silico} ablations.}
    \item \textbf{Introduce the \textit{Rat~v2} body and open-source simulation and training infrastructure.}
\end{enumerate}

Our open-source code and videos are available at: \website
\end{samepage}

\section{Related Work}
\label{sec:2_related_work}

\paragraph{Biomechanical bodies}

Frameworks for embodied simulation now assemble diverse body models across organisms for comparative study~\citep{Arreguit2023FARMSFrameworkAnimal,Zhang2025MIMICMJXNeuromechanicalEmulation}.
For rodent models specifically (i.e. mouse and rat), two prior bodies bracket the relevant abstraction space.
At the detailed end, \citet{Ramalingasetty2024AllFours3D} developed a mouse neuromusculoskeletal body with Hill-type muscles anchored to measured attachment points, afferent feedback pathways (muscle spindle Ia and II, Golgi tendon Ib, and cutaneous afferents), and low-level reflex circuits.
This level of detail enables fine-grained physiological comparisons while making optimization more demanding: their work produced a distribution of fixed-speed trotting solutions, while continuous speed modulation and gait transitions within a single model remained open.
At the abstract end, \citet{Merel2019DeepNeuroethologyVirtual} developed a rat body with anatomically accurate proportions and mass distributions, actuated through joint positions without explicit muscles, and paired with an LSTM controller to support visually guided behaviors including gap crossing and maze foraging.
\citet{Aldarondo2024VirtualRodentPredicts} used this body with MLP and LSTM controllers trained through imitation learning on recorded rat kinematics, yielding neural representations predictive of brain activity.
Both works demonstrated the expressive capacity of joint-level bodies with generic controllers rather than biologically grounded circuit structure, and naturalistic behavior was achieved by imitating recorded kinematics.

Our work builds on the \citet{Merel2019DeepNeuroethologyVirtual} rat body by improving its anatomical fidelity and extending it with a novel flexor-extensor neuromechanical interface that provides an intermediate level of abstraction between joint-level and muscle-level control, enabling systematic comparison of actuation abstractions.
Meanwhile, we view detailed neuromusculoskeletal approaches and our approach as synergistic. An open question is whether the same locomotor phenomena can be modeled across different levels of abstraction, and progress at each level can inform the other.

\paragraph{Neural circuit models}

Locomotion is a useful testbed for biologically grounded control models because the underlying neural circuits are comparatively well characterized, and locomotion requires continuous interactions between neural circuits, the body, and the environment.
There is a long history of biologically inspired locomotion controllers based on central pattern generators~(CPGs)~\citep{Ijspeert2008CentralPatternGenerators,Yu2014SurveyCPGinspiredControl}.
This broad family includes stance-swing transition rules~\citep{Ekeberg2005ComputerSimulationStepping}, parameterized trajectories and footfall templates~\citep{Iscen2019PoliciesModulatingTrajectory}, learned oscillator policies~\citep{Bellegarda2022CPGRLLearningCentral,Shafiee2023PuppeteerMarionetteLearning}, and reduced circuit motifs~\citep{Lodi2018DesignSyntheticCentral}.
Most are biologically inspired rather than tightly constrained in cell types and connectivity.

Both \citet{Ramalingasetty2024AllFours3D} and \citet{Bhattasali2024NeuralCircuitArchitectural} built on cell-type-specific spinal circuit models of mammalian locomotion~\citep{Danner2016CentralControlInterlimb,Danner2017ComputationalModelingSpinal}.
A key difference between them is neuronal abstraction: the former used biophysical neurons with bursting dynamics arising from ionic mechanisms of a persistent sodium current, whereas the latter used simplified neurons with bursting dynamics arising from an abstract relaxation oscillator~\citep{Bhattasali2026SimplifiedModelIntrinsically}.
Also, the former specified more circuitry from biological data, while the latter learned certain modules.

Our work builds on \citet{Bhattasali2024NeuralCircuitArchitectural} with substantial architectural and algorithmic changes that we found were crucial for scaling up to our biorealistic body.
On the architectural side, major additions include a 2-layer afferent feedback module that learns sensory transformations and provides both resetting and non-resetting feedback, and a 3-layer pattern formation module that supports multiple stance/swing subphases, local reflexes, muscle synergies, and trunk actuation.
On the algorithmic side, we rebuild the simulation and training infrastructure from the ground up to support GPU acceleration and modular components for biomechanical and neural circuit models.

\section{Methods}
\label{sec:3_methods}

\subsection{Biomechanical body}
\label{sec:3_biomechanical_body}

\begin{figure}[bt]
\centering
\includegraphics[width=\linewidth]{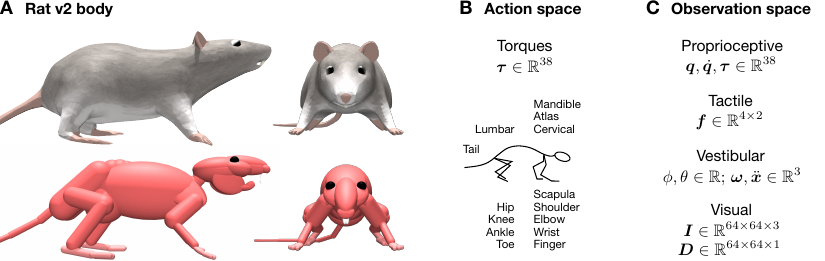}
\caption{
    \textbf{Biomechanical body.}
    \panel{A}~\textit{Rat~v2} has anatomically realistic kinematics, collision geometry, and rendering skins.
    \panel{B}~Actions are torque commands~$\vtau$ for 38 actuated DOFs.
    \panel{C}~Observations are proprioceptive positions~$\vq$, velocities~$\dot{\vq}$, and applied torques~$\vtau$; tactile contact forces~$\vf$; vestibular roll~$\phi$, pitch~$\theta$, angular velocity~$\vomega$, and linear acceleration~$\ddot{\vx}$; visual RGB image~$\mI$ and depth map~$\mD$.
}
\label{fig:1_biomechanical_body}
\vspace{-0.5em}
\end{figure}

We build \textit{Rat~v2}~(\figref[A]{fig:1_biomechanical_body}) from the biomechanical rat body of~\cite{Merel2019DeepNeuroethologyVirtual}, constructed through anatomical measurements of Long-Evans rats (\textit{Rattus norvegicus}) and implemented in the MuJoCo physics engine~\citep{Todorov2012MuJoCoPhysicsEngine}.
We refine the body to improve biological fidelity, extend it with biologically realistic skins to facilitate qualitative behavior evaluation, and integrate it into a new JAX-based environment framework to substantially increase efficiency and configurability.

Actions~$\va_t$ are torque commands~$\vtau$ for 38 actuated degrees of freedom~(DOFs) spanning the head, cervical spine, forelimbs, lumbar spine, hindlimbs, and tail~(\figref[B]{fig:1_biomechanical_body}).
The cervical, lumbar, and tail DOFs use tendons to co-actuate multiple vertebral joints.
Switching the original model from position to torque actuators enables modularly evaluating neuromechanical interfaces~(\autoref{sec:3_neuromechanical_interfaces}).

Observations~$\vo_t$ have 4 modalities~(\figref[C]{fig:1_biomechanical_body}).
Proprioceptive observations provide positions~$\vq$, velocities~$\dot{\vq}$, and applied torques~$\vtau$ for each actuated DOF.
Tactile observations provide contact forces~$\vf$ at the forelimb finger and palm or hindlimb toe and heel.
Vestibular observations provide head roll~$\phi$, pitch~$\theta$, angular velocity~$\vomega$, and linear acceleration~$\ddot{\vx}$.
Visual observations provide an RGB image~$\mI$ and depth map~$\mD$ from a head-mounted camera.

We refine the body specification to improve its ability to reproduce biological behavior, including changes in joint ranges/axes, actuator forces, collision geometry, sensor coverage, default poses, and naming.
For a list of changes and their biological significance, see~\autoref{sec:a1_mujoco_specification}.

We design biologically realistic rendering skins in 3 coat patterns: gray, black-hooded, and cinnamon.
Skins do not affect physics simulation, but they facilitate qualitative visual comparison with experimental observations.
For images and the creation pipeline, see~\autoref{sec:a1_rendering_skins}.

We build an environment framework in JAX~\citep{Bradbury2018JAXComposableTransformations} that yields orders-of-magnitude speedups in parallel simulation, while also enabling modular configuration of observation and action spaces, reward functions, and task logic.
This framework enables large-scale embodied simulation for the neuroscientific community, and we are releasing a separate whitepaper along with the open-source code.
For a software architecture overview, see~\autoref{sec:a1_environment_builder}.

\subsection{Neural circuit model}
\label{sec:3_neural_circuit_model}

\begin{figure}[tb]
\centering
\includegraphics[width=\linewidth]{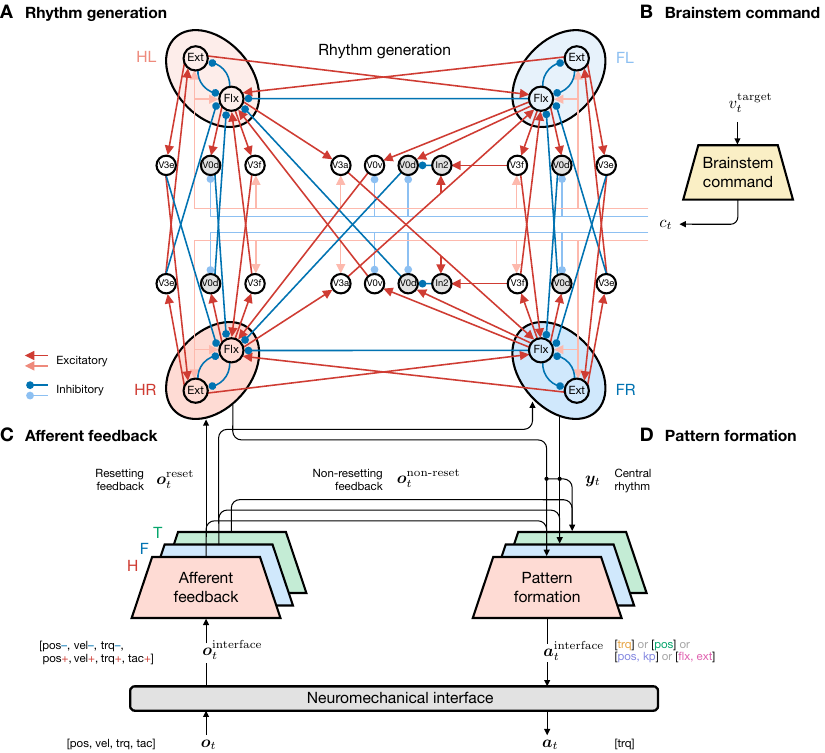}
\caption{
    \textbf{Neural circuit model.}
    \panel{A}~RG organizes limbs into reciprocally inhibiting flexor/extensor half-centers, connected through cross, side, and diagonal commissural interneurons. 
    \panel{B}~BC converts target speed~$v^\text{target}_t$ into brainstem command~$c_t$, modulating half-center frequency/amplitude and interneuron activity in RG to alter gait speed and pattern.
    \panel{C}~AF transforms interface observations~$\vo^\text{interface}_t$ into resetting feedback~$\vo^\text{reset}_t$ and non-resetting feedback~$\vo^\text{non-reset}_t$ passed to RG and PF, respectively.
    \panel{D}~PF combines central rhythm~$\vy_t$ and non-resetting feedback~$\vo^\text{non-reset}_t$ to produce interface actions~$\va^\text{interface}_t$ through delay, synergy, and motor layers.
    Modules are indexed as forelimb~(\textcolor{blue}{F}), hindlimb~(\textcolor{red}{H}), or trunk~(\textcolor{green}{T}), with parameter sharing where possible to exploit body symmetries.
}
\label{fig:2_neural_circuit_model}
\end{figure}

We develop \textit{Rat~NCAP}~(\figref{fig:2_neural_circuit_model}) by extending the neural circuit architectural prior from \cite{Bhattasali2024NeuralCircuitArchitectural}, which builds on spinal circuit models from \cite{Danner2017ComputationalModelingSpinal,Ausborn2019ComputationalModelingBrainstem,Kim2022ContributionAfferentFeedback}.
Scaling to the rat body requires substantial architectural advances as a rat's limb geometry is more complex than a quadruped robot's and requires producing phase-offset joint motions.
Moreover, passive stiffness is not negligible, and trunk joints lie outside the previous 4-limb framework.
We address these challenges by introducing richer RG half-center dynamics, a 2-layer AF module, and a 3-layer PF module with delays, reflexes, and synergies.

The architecture uses 2 neuron types: Basic and Bursting.
The Basic neuron is a standard neuron with internal voltage $v$ that evolves through leaky-integrator dynamics and produces rate-coded activity with a threshold activation function~$f(v) = \text{clip}(v, 0, 1)$.
The Bursting neuron adds an adaptation variable $a$ that interacts with $v$ to produce intrinsic bursting that drives half-center oscillations even without rhythmic input.
For equations and hyperparameters, see~\autoref{sec:a2_neurons}.

The architecture is composed of 4 modules:
Rhythm Generation~(RG) coordinates gaits between limbs, 
Brainstem Command~(BC) provides speed-dependent control of step frequency and gait, 
Afferent Feedback~(AF) transforms sensory observations into half-center and reflex inputs, and 
Pattern Formation~(PF) transforms half-center outputs to motor actions.
\begin{align*}
    c_t &= \text{BrainstemCommand}(v^\text{target}_t) \\
    (\vo^\text{reset}_t, \vo^\text{non-reset}_t) &= \text{AfferentFeedback}(\vo^\text{interface}_t) \\
    (\vy_t, \vh_{t+1}) &= \text{RhythmGeneration}(\vo^\text{reset}_t, \vh_t, c_t) \\
    \va^\text{interface}_t &= \text{PatternFormation}(\vy_t, \vo^\text{non-reset}_t, c_t)
\end{align*}
with neuromechanical interface-specific observations~$\vo^\text{interface}_t$ and actions~$\va^\text{interface}_t$~(\autoref{sec:3_neuromechanical_interfaces}),
task-specific target speed~$v^\text{target}_t$,
brainstem command~$c_t$,
resetting feedback~$\vo^\text{reset}_t$,
non-resetting feedback~$\vo^\text{non-reset}_t$,
central rhythm~$\vy_t$,
and neuron hidden states~$\vh_t$.
Rat~NCAP controls 31 of 38 active DOFs for locomotion (the limbs and trunk), while passive DOFs are held near default poses.
Parameters are indexed by body part~$p \in \{\text{FL, FR, HL, HR, T}\}$ and shared to exploit symmetries.

\begin{figure}[tb]
\centering
\includegraphics[width=\linewidth]{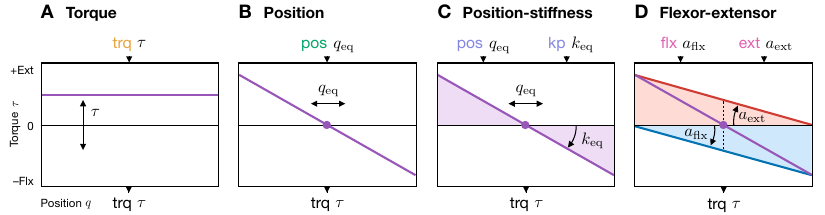}
\caption{
    \textbf{Neuromechanical interfaces.}
    Control laws show applied torque~$\tau$ as a function of the current position~$q$, with control variables reshaping the torque profile.
    \panel{A}~Torque control uses circuit outputs directly as torque~$\tau$ and does not include spring-like negative feedback.
    \panel{B}~Position control only adjusts equilibrium position~$q_\text{eq}$ while using a fixed stiffness.
    \panel{C}~Position-stiffness control explicitly adjusts both equilibrium position~$q_\text{eq}$ and stiffness~$k_\text{eq}$.
    \panel{D}~Flexor-extensor control uses antagonistic muscle activations~$a_\text{flx}, a_\text{ext}$ to implicitly adjust both equilibrium position and stiffness.
}
\vspace{-2em}
\label{fig:3_neuromechanical_interfaces}
\end{figure}

\paragraph{Rhythm generation}
The RG module coordinates limb rhythms to produce speed-dependent gaits.
Each limb is governed by a bursting flexor unit and basic extensor unit that are reciprocally inhibiting.
These limb half-center oscillators are connected through cross, side, and diagonal commissural interneurons, with excitatory synapses promoting synchronization and inhibitory synapses promoting alternation~\citep{Bhattasali2024NeuralCircuitArchitectural}.
To generate larger and stronger steps at higher speeds, we introduce amplitude modulation of half-center activity by BC input in addition to existing frequency modulation, which reflects recruitment of larger neural populations~\citep{Shevtsova2015OrganizationLeftRight} that was not modeled in \cite{Danner2017ComputationalModelingSpinal}.
We tune burst durations to match rat timescales and freeze synaptic weights during training.
For the RG circuit diagram and tuning protocol, see~\autoref{sec:a2_rg}.

\paragraph{Brainstem command}
The BC module produces brainstem command $c_t \in [0, 1]$ to modulate RG half-center frequency/amplitude and interneuron activity.
It uses a linear mapping with hyperparameters~$w^\text{BC}, b^\text{BC}$ to align the command range with the target speed range:
\begin{equation*}
    c_t = f(w^\text{BC} \, v^\text{target}_t + b^\text{BC})
\end{equation*}

\paragraph{Afferent feedback}
The AF module produces sensory feedback for RG and PF through 2 layers.
The sensor layer applies per-element gains and thresholds, modeling tunable sensory receptors:
\begin{equation*}
    \vo^{\text{sensor},p}_t = f(\vw^{\text{sensor},p} \circ \vo^{\text{interface},p}_t + \vb^{\text{sensor},p}), \qquad \vw^{\text{sensor},p} \geq 0, \; \vb^{\text{sensor},p} \leq 0
\end{equation*}
The reset layer produces signed feedback~$\vo^{\text{reset},p}_t \in \R^{\{\text{flx}, \text{ext}\}}$ that resets or entrains the RG rhythm:
\begin{equation*}
    \vo^{\text{reset},p}_t = \mW^{\text{reset},p} \, \vo^{\text{sensor},p}_t + \vb^{\text{reset},p}
\end{equation*}
The sensor outputs also pass directly to PF as non-resetting feedback~$\vo^{\text{non-reset},p}_t = \vo^{\text{sensor},p}_t$, which triggers reflexes without affecting the rhythm.
For AF details, see~\autoref{sec:a2_af}.

\paragraph{Pattern formation}
The PF module produces interface-specific actions from RG through 3 layers.
The delay layer produces copies~\mbox{$k \in \{\text{early}, \text{late}\}$} of each half-center output~\mbox{$h \in \{\text{flx}, \text{ext}\}$}, which have distinct time constants~$T_k \geq 0$ and biases~$b_k \leq 0$ that set their delays:
\begin{equation*}
    z^{h,k,p}_t = \text{Basic}(y^{h,p}_t + b_k;\, T_k)
\end{equation*}
The synergy layer selects synergies by central rhythm, sensory feedback, and brainstem command:
\begin{equation*}
    \vs^p_t = f(\mW^{\text{syn},p} \, \vz^p_t + \mW^{\text{reflex},p} \, \vo^{\text{non-reset},p}_t + \vw^{\text{BC},p} \, c_t)
\end{equation*}
The motor layer maps synergy activations to interface-specific actions:
\begin{equation*}
    \va^{\text{interface},p}_t = f(\mW^{\text{motor},p} \vs^p_t + \vb^{\text{motor},p})
\end{equation*}
The motor weights~$\mW^{\text{motor},p}$ are sign-constrained following biological priors, and biases~$\vb^{\text{motor},p}$ are initialized to produce the default standing pose.
For PF details, see~\autoref{sec:a2_pf}.

\subsection{Neuromechanical interfaces}
\label{sec:3_neuromechanical_interfaces}

We design 4 neuromechanical interfaces to explore how biorealistic circuits may benefit from biorealistic actuation.
All interfaces share an observation transform from body observations~$\vo_t$ to interface observations~$\vo^\text{interface}_t$, rectifying proprioceptive and tactile signals into positive and negative components like afferents.
The interfaces differ in the action transform from interface actions~$\va^\text{interface}_t$ to body actions~$\va_t$, applying a per-DOF control law with control variables output by the circuit.

Torque control~(\figref[A]{fig:3_neuromechanical_interfaces}) sets torques directly:\enspace
$\tau = \textcolor{yellow}{\tau}$

Position control~(\figref[B]{fig:3_neuromechanical_interfaces}) adjusts only equilibrium position with a fixed stiffness:\enspace
$\tau = k_\text{eq}\cdot(\textcolor{green}{q_\text{eq}} - q)$

Position-stiffness control~(\figref[C]{fig:3_neuromechanical_interfaces}) explicitly adjusts both equilibrium position and stiffness, which biological limbs regulate to stabilize posture and movement~\citep{Shadmehr1993ControlEquilibriumPosition}:\enspace
$\tau = \textcolor{indigo}{k_\text{eq}}\cdot(\textcolor{indigo}{q_\text{eq}} - q)$

Flexor-extensor control~(\figref[D]{fig:3_neuromechanical_interfaces}) uses antagonistic muscles to implicitly adjust both equilibrium position and stiffness: differential activation shifts the equilibrium, and co-activation scales stiffness.
For the derivation and biological motivation, see~\autoref{sec:a3_interfaces}.
The circuit controls flexor and extensor muscle activations:\enspace
\mbox{$\tau = \textcolor{magenta}{a_\text{ext}}\cdot k_\text{ext}\cdot(q_\text{ext} - q) \;+\; \textcolor{magenta}{a_\text{flx}}\cdot k_\text{flx}\cdot(q_\text{flx} - q)$}

\subsection{Training setup}
\label{sec:3_training_setup}

We jointly train a single model on flat-ground locomotion with 6 subtasks, each tracking a fixed target speed~\mbox{($v^\text{target} \in \{0.0, 0.2, 0.4, 0.6, 0.8, 1.0\}$~m/s)}, using velocity-tracking rewards and a pose reward for standing.
We use Augmented~Random~Search~(ARS)~\citep{Mania2018SimpleRandomSearch} to optimize AF and PF parameters, which avoids backpropagation through neural dynamics and the simulator and is highly parallelizable on a GPU.
For task, reward, and algorithm details, see~\autoref{sec:a4_training}.

\section{Results}
\label{sec:4_results}

\subsection{Rat NCAP produces variable-speed locomotion and gait transitions}

\begin{figure}[b]
\vspace{-0.5em}
\centering
\includegraphics[width=\linewidth]{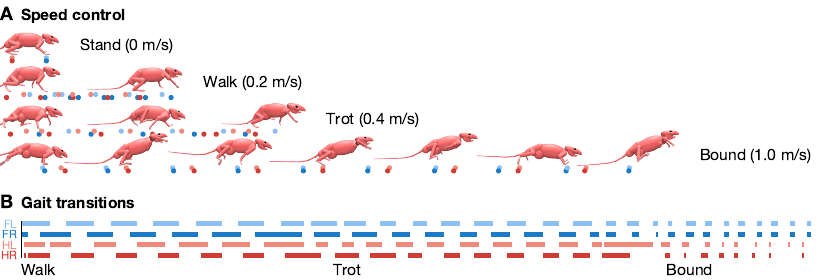}
\caption{
    \textbf{Locomotion behavior.}
    \panel{A}~Timelapse frames and footsteps over 2~s at each target speed.
    \panel{B}~Footfall pattern as brainstem command increases within an episode to produce gait transitions.
}
\label{fig:4_locomotion_behavior}
\end{figure}

\begin{figure}[t]
\centering
\includegraphics[width=\linewidth]{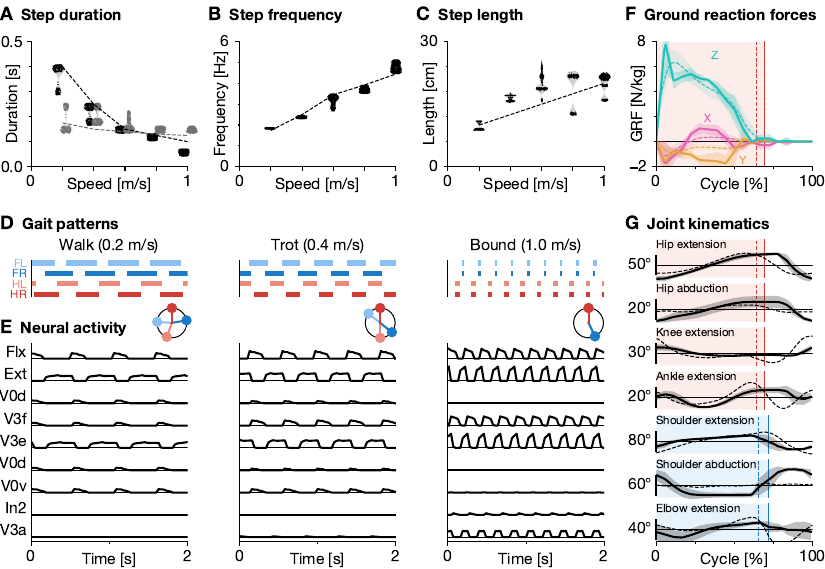}
\caption{
    \textbf{Experimental validation.}
    \panel{A--C}~Step statistics across target speeds: stance~(\textcolor{black}{black}) and swing~(\textcolor{gray}{gray}) durations, step frequency, and step length.
    Speed increases mainly through shorter stance, as in rats.
    \panel{D}~Footfall patterns and interlimb phases for representative walk, trot, and bound.
    Gaits shift from left-right alternation to synchrony as speed increases.
    \panel{E}~RG cell-type activity for the same gaits.
    V0d and V0v activity decline and V3 activity rises with speed.
    \panel{F}~Ground reaction forces of the \textcolor{red}{HR}~limb: forward~\textcolor{magenta}{X}, lateral~\textcolor{yellow}{Y}, and vertical~\textcolor{cyan}{Z}, with stance shaded and liftoff marked.
    Forward force shows braking then propulsion.
    \panel{G}~Joint kinematics of the \textcolor{red}{HR} and \textcolor{blue}{FR} limbs.
    Model data over 3 seeds, mean~$\pm$~SD.
    Rat data~(dashed) from (\textbf{A})~\citet{Gillis2001HindlimbMuscleFunction}, (\textbf{B},~\textbf{C})~\citet{Clarke1986QuantitativeStudyNormal}, (\textbf{F},~\textbf{G}~hind)~\citet{Dienes2022ComprehensiveDynamicKinematic}, and (\textbf{G}~fore)~\citet{Bonnan2016ForelimbKinematicsRats}.
}
\label{fig:5_experimental_validation}
\vspace{-0.5em}
\end{figure}

Rat~NCAP learns to stand, walk, trot, and bound in a single model, with naturalistic, speed-dependent gaits~(\figref[A]{fig:4_locomotion_behavior}, \webref{Video~1}).
Although trained only at fixed speeds, Rat~NCAP transitions from walk to trot to bound as brainstem command increases mid-episode~(\figref[B]{fig:4_locomotion_behavior}, \webref{Video~2}).
Prior work in this rat relied on imitation learning for naturalistic behavior~\citep{Aldarondo2024VirtualRodentPredicts,Zhang2025MIMICMJXNeuromechanicalEmulation} or produced less naturalistic movement without it~\citep{Merel2019DeepNeuroethologyVirtual}.
Rat~NCAP uses only simple rewards, suggesting that its biologically grounded architecture provides an effective inductive bias.

Comparisons with an MLP and Quadruped~NCAP~\citep{Bhattasali2024NeuralCircuitArchitectural} support this interpretation.
Rat~NCAP achieves return comparable to or higher than MLPs trained with deep reinforcement learning or the same ARS algorithm~(\figref{fig:a6_mlp_training}), yet both MLPs produce unnatural poses and gaits~(\figref{fig:a6_mlp_behavior}).
Rat~NCAP also generalizes at least as well to unseen bumpy terrain~(\figref{fig:a6_mlp_generalization}).
Finally, it achieves dramatically higher return than Quadruped~NCAP, demonstrating that its architectural additions are critical for scaling to the more complex, biorealistic body~(\figref{fig:a6_quadruped_baseline}).

\subsection{Trained behavior matches experimental data on rat locomotion}

Producing forward locomotion is necessary but not sufficient for a biologically grounded model, so we compare the trained behavior and neural activity with experimental data.
Rat~NCAP modulates speed as rats do: stance shortens while swing remains roughly constant~(\figref[A]{fig:5_experimental_validation}), raising step frequency within the experimentally reported range~(\figref[B]{fig:5_experimental_validation}).
This asymmetry is consistent with flexor-dominated rhythm generation, in which flexor bursts driving swing remain brief while extensor bursts driving stance shorten with speed~(\figref[E]{fig:5_experimental_validation}).
Step length matches reported values at the slowest and fastest speeds but exceeds them at intermediate speeds~(\figref[C]{fig:5_experimental_validation}).

Gait patterns change with speed, as interlimb phases shift from alternation in walk and trot to synchrony in bound~(\figref[D]{fig:5_experimental_validation}).
RG activity shifts accordingly~(\figref[E]{fig:5_experimental_validation}): from walk to trot, V0d activity declines and V0v activity sustains diagonal alternation; in bound, both vanish as V3 activity promotes left-right synchrony, matching data-constrained biophysical models~\citep{Zhang2022RoleV3Neurons}.

Limb kinetics and kinematics are also consistent with experimental data.
Vertical force peaks early in stance, forward force shows braking then propulsion~(\figref[F]{fig:5_experimental_validation}), and joint angles follow experimental profiles, including phase offsets between proximal and distal joints~(\figref[G]{fig:5_experimental_validation}).
Agreement is closest during stance; during swing, the knee and elbow flex less than in rats, which may reflect limitations of the circuit or body model.
Constraining non-planar limb DOFs substantially lowers return, whereas constraining lumbar DOFs has less effect, likely due to compensation~(\figref{fig:a7_dof_ablation}), though they are recruited in a speed-dependent manner when available~(\figref{fig:a7_lumbar_kinematics}).
Together, these results indicate that biologically grounded architecture shapes features of locomotion beyond the training rewards.

\subsection{Neuromechanical interfaces strongly influence trainability and gait quality}

\begin{figure}[tb]
\centering
\includegraphics[width=\linewidth]{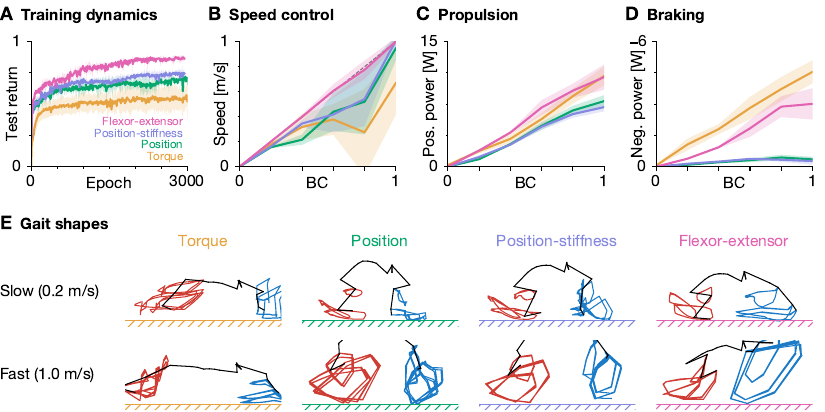}
\caption{
    \textbf{Neuromechanical interface comparison.}
    \panel{A}~Test return over training epochs for all 4 interfaces.
    Flexor-extensor achieves the highest return, and torque the lowest.
    \panel{B}~Speed as brainstem command increases from 0~to~1.
    Flexor-extensor tracks target speed most closely, while torque fails at high commands.
    \panel{C, D}~Positive~(propulsion) and negative~(braking) mechanical power, summed over actuators and averaged over timesteps.
    Flexor-extensor matches torque in propulsion and absorbs more energy than position-based interfaces, consistent with spring-like compliance.
    \panel{E}~Sagittal trajectories of the \textcolor{red}{HR} and \textcolor{blue}{FR} limbs at slow and fast speeds.
    Flexor-extensor produces rounded trajectories, while position produces sharp arcs.
    Model data over 3 seeds, mean~$\pm$~SD.
}
\label{fig:6_neuromechanical_comparison}
\vspace{-0.5em}
\end{figure}

We test whether a biologically grounded neural circuit benefits from biorealistic actuation by training Rat~NCAP with each neuromechanical interface.
Flexor-extensor control achieves the highest test return and tracks target speed most accurately, followed by position-stiffness and position control, while torque control achieves the lowest return with this architecture~(\figref[A, B]{fig:6_neuromechanical_comparison}).

Flexor-extensor control also produces qualitatively different gaits from the common choice of position control.
It generates greater propulsion~(\figref[C]{fig:6_neuromechanical_comparison}) and absorbs substantially more energy~(\figref[D]{fig:6_neuromechanical_comparison}), as its antagonistic springs store and return energy analogous to tendon compliance.
Correspondingly, its limb trajectories are rounded, with extended stance as the paw follows the ground, whereas position control produces sharp arcs with brief ground contact~(\figref[E]{fig:6_neuromechanical_comparison}, \webref{Video~3}).
Torque control absorbs the most energy, though its disorganized trajectories suggest uncoordinated braking rather than compliance.
The benefits of flexor-extensor control are consistent with the view that animals act through muscles whose intrinsic mechanics provide stabilizing feedback rather than commanding joint torques~\citep{Dickinson2000HowAnimalsMove}.
Embodied models of motor circuits may thus benefit from muscle-like actuation even in this simple form, at little added complexity over position control.

\subsection{Cell-type ablations reproduce known phenotypes and produce testable predictions}

\begin{figure}[tb]
\centering
\includegraphics[width=\linewidth]{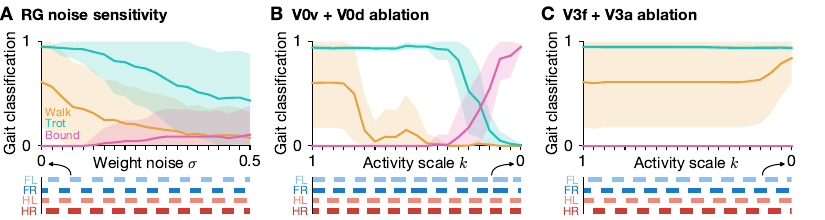}
\caption{
    \textbf{Cell-type ablations.}
    Fraction of step cycles classified as each gait by idealized interlimb phase ranges at 0.4~m/s, with footfalls for the indicated condition.
    \panel{A}~RG weights scaled by multiplicative Gaussian noise of standard deviation~$\sigma$.
    Trot remains dominant at all noise levels.
    \panel{B}~V0v and V0d activity scaled by factor~$k$.
    Trot switches abruptly to bound, reproducing the knockout phenotype at $k = 0$.
    \panel{C}~V3f and V3a activity scaled by factor~$k$.
    Trot persists, so not all populations are equally critical.
    Model data over 3 seeds and 30 trials, mean~$\pm$~SD.
}
\label{fig:7_cell_type_ablations}
\vspace{-0.5em}
\end{figure}

A central advantage of biologically grounded circuit models is the ability to ablate identified cell types \textit{in silico}.
Since training does not include RG ablations and RG remains fixed, these experiments test whether phenotypes of the non-embodied RG model persist when it drives a body through learned AF and PF modules.
As in other biologically grounded and connectome-constrained models, RG synaptic strengths are not identifiable from experimental data, so we first test sensitivity to them.
Under sign-preserving multiplicative noise, trot degrades only gradually and remains the dominant gait even at $\sigma = 0.5$~(\figref[A]{fig:7_cell_type_ablations}), indicating that the gaits do not depend on finely tuned weights.

In mice, ablating V0v neurons eliminates trot, and additionally ablating V0d neurons leaves bound as the only coordinated gait~\citep{Bellardita2015PhenotypicCharacterizationSpeedAssociated}.
Rat~NCAP reproduces this combined phenotype: scaling V0v and V0d activity to zero replaces trot with bound~(\figref[B]{fig:7_cell_type_ablations}, \webref{Video~4}).
Removing V0-mediated alternation likely leaves V3-mediated synchrony unopposed, so the remaining commissural drive produces bound.
Graded scaling further predicts a threshold-like transition, as trot persists until $k \approx 0.35$ and then gives way to bound by $k \approx 0.1$, consistent with abrupt gait changes in biophysical models as commissural drive varies~\citep{Danner2017ComputationalModelingSpinal}.
Such graded reductions could be tested experimentally by partially reducing cell counts, intrinsic excitability, or synaptic efficacy in these populations.
In contrast, scaling V3f and V3a activity to zero leaves trot intact~(\figref[C]{fig:7_cell_type_ablations}), so not all commissural populations are equally critical for trot.
Together, these results show that RG cell-type phenotypes persist in an embodied setting and yield experimentally testable predictions.

\section{Discussion}
\label{sec:5_discussion}

This work asked how to develop embodied simulations that enable the study of increasingly complex bodies and neural circuits.
Our results support an intermediate level of abstraction: a biorealistic rat body controlled by a biologically grounded architecture and trained with simple rewards produces naturalistic locomotion and enables circuit-level analyses.
A cell-type-specific spinal circuit model, previously studied in non-embodied settings, retains its gait-generating function when coupled to realistic body dynamics and closed-loop feedback.
The resulting model supports validation against behavioral and neural data, reveals benefits of muscle-like actuation for circuit function, and predicts responses to perturbations not yet tested experimentally.
This balance is practically useful for building embodied simulations and scientifically useful for evaluating circuit hypotheses in complex animals.

Our work has several limitations.
First, behavior does not yet match rats in all respects, including step length at intermediate speeds and distal joint kinematics during swing.
These discrepancies may arise from the circuit model, the neuromechanical interface, or the body, including the difficulty of representing compliance in rigid-body physics engines.
Comparing the behavior of Rat~NCAP with that of imitation-learned policies on the same body could separate these factors, since such policies approximate rat behavior within the kinematic limits of Rat~v2.
Second, circuit parameters are underconstrained: RG weights are not identifiable from available data, and the learned AF and PF modules find plausible rather than unique solutions.
How much these connections vary across animals is unknown; if they vary little, future work could identify the tasks or rewards under which they reliably emerge, and if they vary widely, how developmental plasticity organizes them.

Our work can also be extended in several directions.
Actuation could be refined by augmenting the flexor-extensor interface with force-velocity relationships and activation dynamics.
Additional spinal and brainstem circuits could enable new behaviors, such as turning, postural adjustments, and non-periodic movements like rearing, jumping, and grooming.
More complex environments with uneven terrain and perturbation forces could test robustness beyond flat-ground locomotion.
A higher-level controller could sequence the resulting motor primitives to perform cognitive tasks such as navigation and foraging.
Together, this work lays a foundation for embodied simulations that move from spinal locomotion toward naturalistic whole-body behavior in increasingly complex animals, while yielding mechanistic insights into the neural circuits that generate it.

\begin{ack}
Thanks to Charles~Zhang and Josh~Merel for assistance with the biomechanical body model.
Thanks to Jonathan~Arreguit, Elliot~Abe, Eric~Leonardis, Simon~Danner, Shravan~Ramalingasetty, Gizem~Özdil, Pavel~Tolmachev, and many others for insightful discussions and feedback.

This work was supported by the Fannie and John Hertz Foundation Fellowship~(NXB);
the Packard Fellowship for Science and Engineering~(LP);
the Institute of Information \& Communications Technology Planning \& Evaluation~(IITP) with a grant by the Ministry of Science and ICT~(MSIT) of the Republic of Korea in connection with the Global AI Frontier Lab International Collaborative Research Agreement (RS-2024-00469482, RS-2024-00509279)~(LP);
and the National Science Foundation AI Research Institute on Interaction for AI Assistants~(ARIA) (No.~2433429)~(JC,~GWL).

\section*{Author Contributions}
Conceptualization (NXB, JC, GWL). Investigation (NXB, JC). Software (NXB, JC). Supervision (GWL, LP). Writing: original draft (NXB, JC), review and editing (NXB, JC, GWL, LP).

\end{ack}

{
  \small
  \bibliographystyle{abbrvnat}
  \bibliography{bibliography.bib}
}
\FloatBarrier
\newpage

\appendix
\counterwithin{figure}{section}
\renewcommand{\floatpagefraction}{0.5}
\section{Biomechanical Body}
\label{sec:a1_biomechanical}

\subsection{MuJoCo specification}
\label{sec:a1_mujoco_specification}

The MuJoCo XML file specifies the morphology, dynamics, actuators, and sensors of the body.
We made the following changes to the body of~\citet{Merel2019DeepNeuroethologyVirtual} to improve fidelity and usability:
\begin{itemize}[leftmargin=0.2in, nosep, parsep=0.5ex]
    \item \textbf{Joint axes:} Shoulder supination, lumbar, cervical, and tail joints are flipped so that positive commands correspond to outward rotation, rightward bending, and upward extension. Toe and finger joints are flipped so that positive commands correspond to extension.
    \item \textbf{Joint ranges:} Asymmetric limits are applied to joints with asymmetric anatomical ranges, symmetric limits are recalibrated, and toe and finger ranges are expanded to permit flexion.
    \item \textbf{Default pose:} The default pose is a neutral standing pose with closed mouth.
    \item \textbf{Collision geometry:} Limb geometry is revised so that all 4 paws rest on the ground in the default pose, and the tail is simplified from 28 to 6 bodies to reduce computation.
    \item \textbf{Touch sensors:} Each paw has 2 sensors~(digit and palm/heel) that report contact force rather than binary contact, providing graded information about paw loading.
    \item \textbf{Actuators:} Force limits are enabled, force ranges are recalibrated from stiffness and joint limits, and joint and tendon actuators are unified into a single interface.
    \item \textbf{Conventions:} Body parts use anatomical names~(e.g., ``caudal'' $\to$ ``tail''), units are degrees rather than radians, and actuators are ordered rostral-to-caudal, left-to-right, and proximal-to-distal.
    \item \textbf{Sensors and cameras:} An orientation sensor and 5 cameras support analysis and rendering.
\end{itemize}
Motion-capture keypoints are preserved, so that Rat~v2 remains compatible with existing imitation datasets.
The resulting action and observation spaces are listed in~\autoref{tab:a1_spaces}.

\begin{table}[h]
\caption{
    \textbf{Action and observation spaces.}
    Actions by body segment and joint~(left) and observations by modality~(right), with dimensions in brackets.
    Limb joints have left and right copies.
}
\label{tab:a1_spaces}
\small
\setlength{\tabcolsep}{4pt}
\noindent
\begin{tabular}[t]{@{}llr@{}}
\toprule
Segment & Action & Dim. \\
\midrule
Head & Atlas: extend & [1] \\
 & Mandible: open & [1] \\
Neck & Cervical: extend, bend, twist & [3] \\
Forelimb & Scapula: supinate, abduct, extend & [2, 3] \\
 & Shoulder: supinate, extend & [2, 2] \\
 & Elbow: extend & [2, 1] \\
 & Wrist: extend & [2, 1] \\
 & Finger: extend & [2, 1] \\
Trunk & Lumbar: extend, bend, twist & [3] \\
Hindlimb & Hip: supinate, abduct, extend & [2, 3] \\
 & Knee: extend & [2, 1] \\
 & Ankle: extend & [2, 1] \\
 & Toe: extend & [2, 1] \\
Tail & Caudal: extend, bend & [2] \\
\midrule
\textbf{Total} & & \textbf{[38]} \\
\bottomrule
\end{tabular}%
\hfill
\begin{tabular}[t]{@{}llr@{}}
\toprule
Modality & Observation & Dim. \\
\midrule
Proprioceptive & Positions~$\vq$ & [38] \\
 & Velocities~$\dot{\vq}$ & [38] \\
 & Torques~$\vtau$ & [38] \\
Tactile & Contact forces~$\vf$ & [4, 2] \\
Vestibular & Roll~$\phi$ & [1] \\
 & Pitch~$\theta$ & [1] \\
 & Angular velocity~$\vomega$ & [3] \\
 & Linear acceleration~$\ddot{\vx}$ & [3] \\
Visual & RGB image~$\mI$ & [64, 64, 3] \\
 & Depth map~$\mD$ & [64, 64, 1] \\
\bottomrule
\end{tabular}
\end{table}

\clearpage
\subsection{Rendering skins}
\label{sec:a1_rendering_skins}

\begin{figure}[h]
\centering
\includegraphics[width=\linewidth]{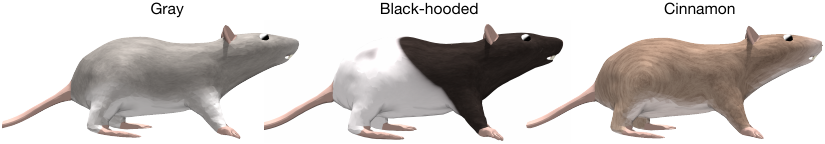}
\caption{
    \textbf{Rendering skins.}
    Renderings of the 3 Long-Evans coat patterns.
    Skins facilitate qualitative comparison with experimental videos without affecting physics.
}
\label{fig:a1_rat_skins}
\end{figure}

We designed biologically realistic rendering skins in Blender to enable qualitative comparison between simulated and experimentally observed rat locomotion~(\autoref{fig:a1_rat_skins}).
The MuJoCo body was imported into Blender, where each joint and body segment became a bone in an armature.
A low-polygon mesh was built from reference images of Long-Evans rats, mirrored to enforce bilateral symmetry, and refined with hand-added asymmetries~(e.g., tail curvature).
The mesh was subdivided into a high-polygon surface, and bone weights were painted onto vertices to determine mesh deformation.
Procedural textures model 3 Long-Evans coat patterns: gray, black-hooded, and cinnamon.
A custom Blender plugin exports the rigged mesh and texture to MuJoCo's skin~(\texttt{.skn}) and texture~(\texttt{.png}) formats.
Skins are purely cosmetic and do not affect the physics simulation.

\subsection{Environment framework}
\label{sec:a1_environment_builder}

\begin{figure}[h]
\centering
\includegraphics{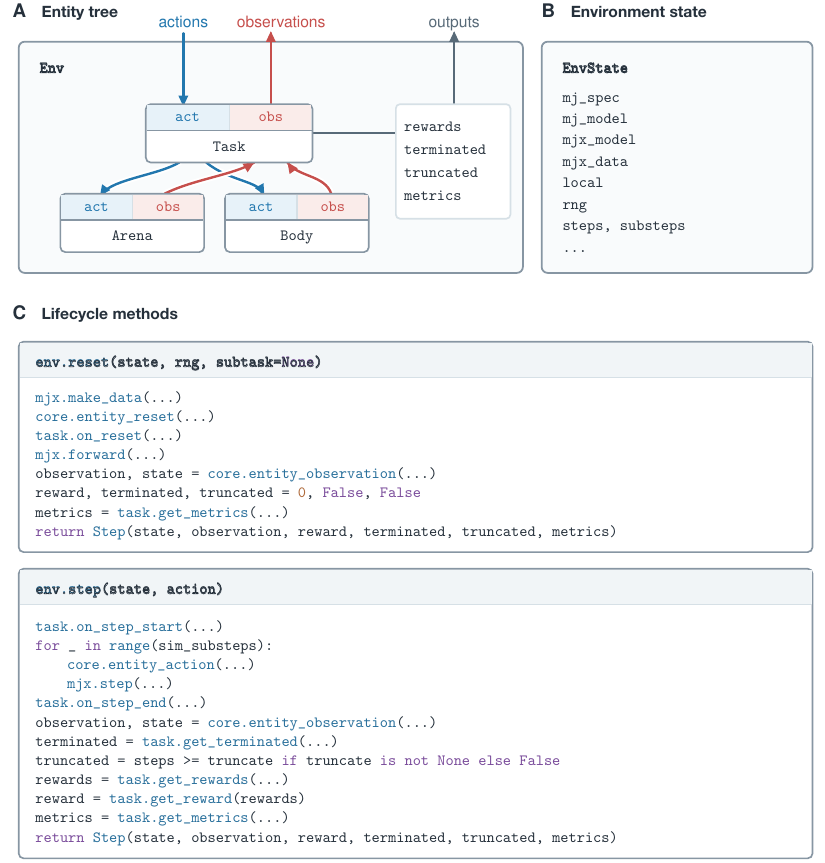}
\caption{
    \textbf{Environment framework.}
    \panel{A}~\texttt{Entity} tree, shown for a \texttt{Task} with arena and body entities.
    Actions pass down and observations pass up through each entity's \texttt{Transform} objects, which can be swapped without modifying other entities.
    \panel{B}~Environment state.
    All simulation state is held in an immutable, explicitly passed \texttt{EnvState}, so environments compile and vectorize in parallel.
    \panel{C}~Lifecycle methods.
    \texttt{reset} and \texttt{step} call \texttt{Task} and \texttt{Entity} hooks around physics updates.
}
\label{fig:a1_framework}
\end{figure}

Training large populations of neural circuit models requires efficient, high-throughput simulation.
We build an environment framework in JAX that adapts the \texttt{dm\_control} composer~\citep{Tassa2020DmControlSoftware} to XLA compilation and GPU-accelerated physics~(\autoref{fig:a1_framework}).
The framework separates environment composition, simulation state, and execution, so that bodies, arenas, interfaces, and tasks can be reconfigured across experiments without changing the simulation code.

\paragraph{Entity tree}
An \texttt{Env} is composed of a tree of \texttt{Entity} objects with a \texttt{Task} at its root~(\figref[A]{fig:a1_framework}).
Each \texttt{Entity} contributes its own model elements, such as bodies, sensors, actuators, or terrain, and the \texttt{Task} is a special \texttt{Entity} that additionally defines rewards, termination and truncation conditions, metrics, and subtasks.
Each \texttt{Entity} has an action \texttt{Transform} that maps actions received from its parent to its own actions, and an observation \texttt{Transform} that maps its own observations to those exposed to its parent.
\texttt{Transform} objects make components interchangeable: each neuromechanical interface is a pair of body \texttt{Transform} objects, swappable without modifying the \texttt{Task} or arena.

\paragraph{Environment state}
All simulation state is stored in an immutable \texttt{EnvState}~(\figref[B]{fig:a1_framework}), including the MuJoCo model and data, entity-local state, the random number generator key, and step counters.
\texttt{Env} methods receive and return this state explicitly rather than modifying it in place, so that they can be compiled with XLA and vectorized across thousands of parallel environments.
\texttt{Entity} and \texttt{EnvState} objects are defined with the Equinox library~\citep{Kidger2021EquinoxNeuralNetworks}.

\paragraph{Lifecycle methods}
The \texttt{Env} exposes \texttt{reset} and \texttt{step} methods~(\figref[C]{fig:a1_framework}).
\texttt{reset} initializes the physics data, resets each \texttt{Entity}, and computes initial observations.
\texttt{step} applies action \texttt{Transform} objects and advances physics for each substep, then computes observations, rewards, termination, truncation, and metrics.
\texttt{Task} hooks such as \texttt{on\_reset}, \texttt{on\_step\_start}, and \texttt{on\_step\_end} implement task-specific behavior, such as sampling target speeds, without modifying the physics loop.

\paragraph{Physics backends}
Physics runs through MuJoCo XLA with either its XLA or Warp backend, so the same environment code runs on CPU, GPU, and TPU.
The full simulation step, including contact detection, constraint solving, and sensor computation, compiles into GPU kernels, yielding large speedups over serial CPU simulation.
Experiments in this paper use the Warp backend.

\clearpage
\section{Neural Circuit Model}
\label{sec:a2_neural_circuit}

\subsection{Neurons}
\label{sec:a2_neurons}

The architecture uses 2 rate-coded neuron types with excitatory or inhibitory synaptic input.

\paragraph{Basic neuron}
The Basic neuron is a rate-coded leaky integrator with time constant~$T_v$, bias~$B$, weighted input~$x$, internal voltage~$v$, and output activity~$y$:
\begin{gather*}
    x = \text{clip}\!\left( B + \textstyle\sum_i w_i y_i,\, -1, 1 \right), \qquad
    \frac{T_v}{4} \frac{dv}{dt} = x - v, \qquad
    y = f(v) = \text{clip}(v, 0, 1)
\end{gather*}
The neuron produces activity proportional to its voltage once the voltage exceeds zero.

\paragraph{Bursting neuron}
The Bursting neuron is a relaxation oscillator that produces intrinsic bursting dynamics through 2 coupled variables: voltage~$v \in [-1, 1]$ and adaptation~$a \in [0, 1]$.
Adaptation depletes during the active phase and recovers during the quiet phase, driving alternating transitions between states.
Input modulates the active and quiet durations, enabling phase resetting and entrainment.
For equations and analysis, see~\citet{Bhattasali2026SimplifiedModelIntrinsically}.

\subsection{Rhythm generation}
\label{sec:a2_rg}

\begin{figure}[p]
\centering
\includegraphics[width=\linewidth]{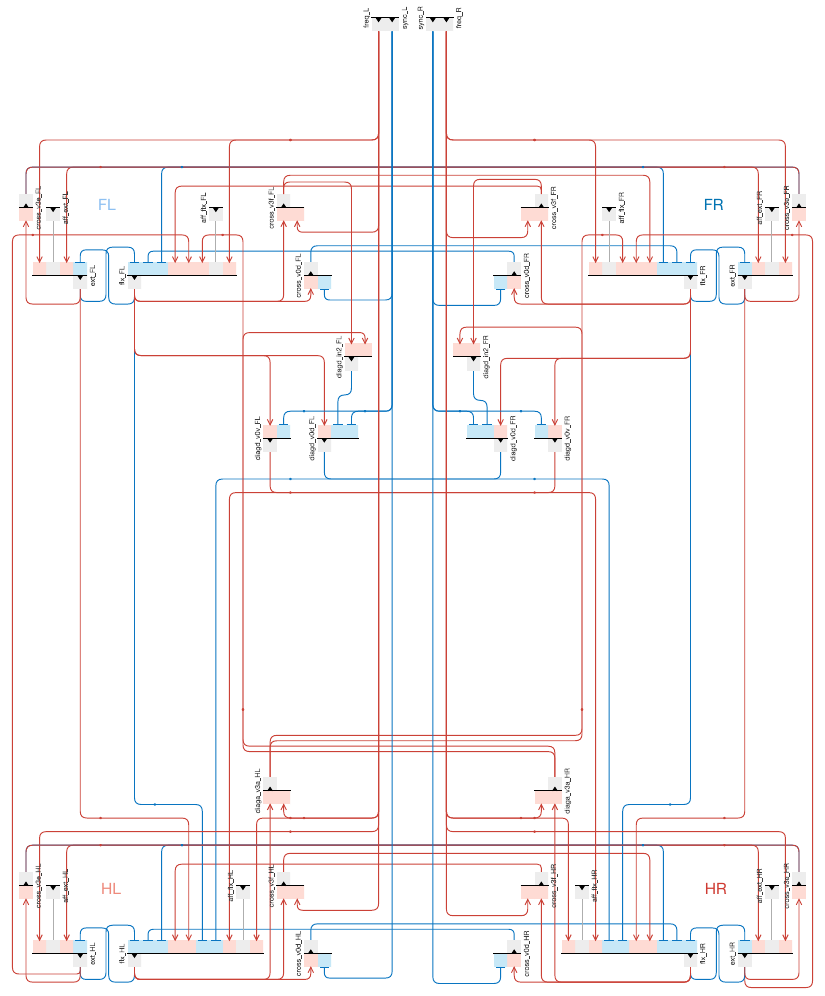}
\caption{
    \textbf{RG circuit.}
    Full circuit schematic showing cell-type-specific connectivity between limb half-centers and interneuron populations.
    Excitatory synapses are shown in \textcolor{red}{red}, and inhibitory synapses in \textcolor{blue}{blue}.
    Commissural interneurons~(V0d, V0v, V3) mediate left-right coordination, and long propriospinal neurons mediate forelimb-hindlimb coordination.
}
\label{fig:a2_rg_circuit}
\end{figure}

RG follows the cell types, neurotransmitter identities, and connectivity of the spinal circuit model of~\citet{Danner2017ComputationalModelingSpinal}~(\autoref{fig:a2_rg_circuit}).
Each limb has a Bursting flexor and a Basic extensor half-center.
Commissural interneurons~(V0d, V0v, V3) couple left and right limbs, and long propriospinal neurons couple forelimbs and hindlimbs through diagonal and homolateral pathways.

\paragraph{Tuning protocol}
RG weights and Bursting neuron hyperparameters~(time constants, thresholds) were hand-tuned to produce the gait transition sequence stand~$\to$ walk~$\to$ trot~$\to$ bound as brainstem command~$c_t$ increases from 0 to 1.
Tuning followed the protocol of \citet{Bhattasali2024NeuralCircuitArchitectural}, using its weights as a starting point and adjusting time constants to match rat stepping, which is faster than that of a quadruped robot.
RG weights are frozen throughout training.

\begin{figure}[p]
\centering
\includegraphics[width=\linewidth]{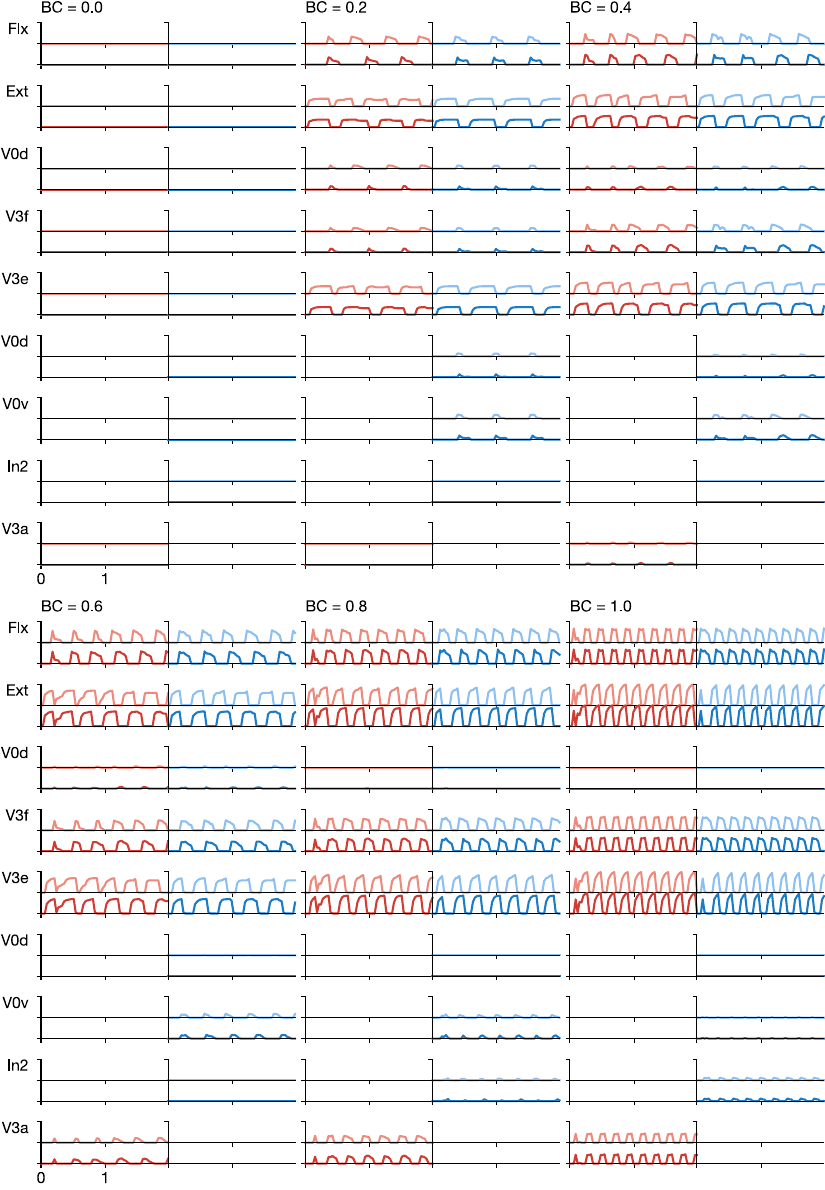}
\caption{
    \textbf{RG activity.}
    Activity of RG half-centers and interneuron populations over 2~s at increasing brainstem command~(BC), for \textcolor{red}{hindlimbs} and \textcolor{blue}{forelimbs}, with left and right sides in light and dark traces.
    The first V0d row shows commissural V0d neurons, and the second shows diagonal V0d neurons.
    As BC increases, stepping frequency increases, V0d and V0v activity decline, and V3 activity increases, matching speed-dependent recruitment in biophysical models~\citep{Zhang2022RoleV3Neurons}.
}
\label{fig:a2_rg_activity}
\end{figure}

\paragraph{Cell-type activity}
RG activity reproduces key features of speed-dependent cell-type activity reported for biophysical models validated against experimental data~(\autoref{fig:a2_rg_activity}; compare Figure~10 of~\citealp{Zhang2022RoleV3Neurons}).
Inhibitory V0d activity is synchronous with flexor bursts and decreases toward high speeds.
Excitatory V0v activity also decreases toward high speeds, but more slowly than V0d activity, which supports trot at intermediate speeds.
Diagonal V3 activity increases toward high speeds as bound emerges.
Half-center oscillations are flexor-dominated, with rhythm arising from the intrinsic bursting of the flexor half-center rather than from escape of the extensor half-center.

\subsection{Brainstem command}
\label{sec:a2_bc}

BC maps target speed to brainstem command through a fixed linear mapping, $c_t = f(v^\text{target}_t)$, so that the command equals the target speed over the trained range.
The command modulates RG half-center frequency and amplitude, interneuron activity, and PF synergies.
BC has no trainable parameters.

\subsection{Afferent feedback}
\label{sec:a2_af}

\paragraph{Modules}
AF consists of forelimb~(F), hindlimb~(H), and trunk~(T) modules.
The F module is applied to FL and FR, and the H module to HL and HR, so that weights are shared between left and right limbs.
Limb modules observe joint positions, velocities, and torques of their DOFs and the 2 tactile sensors of their paw.
The T module observes joint positions, velocities, and torques of the lumbar DOFs, with bending and twisting observations mirrored into left and right copies.
Each proprioceptive signal is rectified into positive and negative components~(\autoref{sec:a3_observations}).

\paragraph{Sensor layer}
The sensor layer applies a per-element gain and threshold followed by~$f$, modeling the saturating response of sensory receptors.
Sensor-layer outputs~$\vo^{\text{sensor},p}_t$ pass to the reset layer and to PF as non-resetting feedback, which triggers reflexes without affecting the rhythm.

\paragraph{Reset layer}
The reset layer maps each limb's sensor-layer outputs to resetting feedback for its own flexor and extensor half-centers, followed by clipping to $[-1, 1]$.
Weights from other limbs and from the T module are fixed at 0, so the T module acts only through PF.

\begin{figure}[h]
\centering
\includegraphics{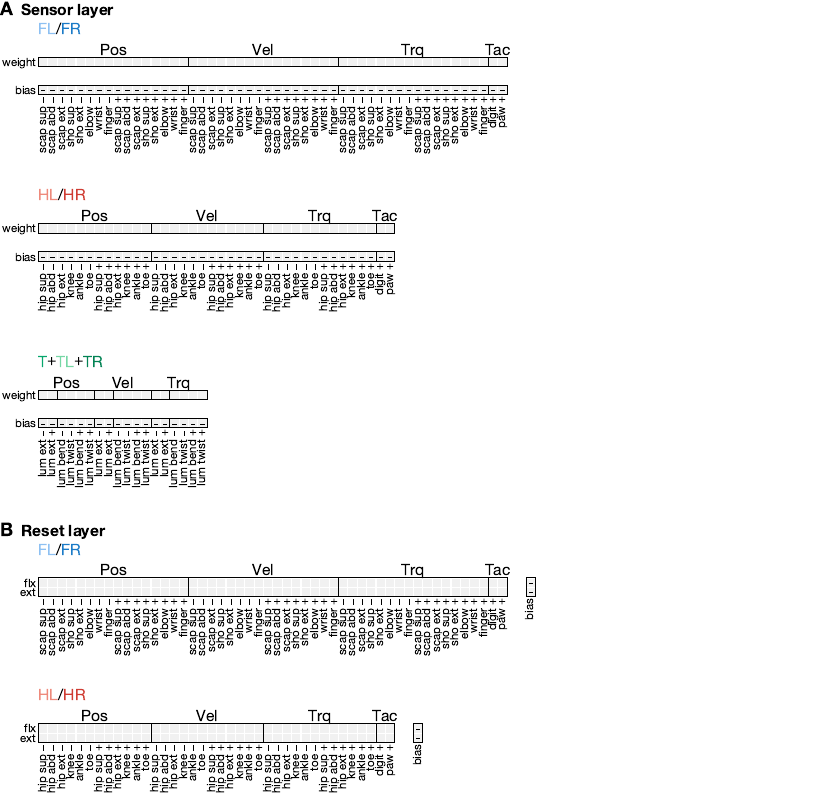}
\caption{
    \textbf{AF initialization.}
    \panel{A}~Sensor-layer weights and biases of the F and H modules.
    Biases are constrained to be nonpositive, so that each afferent responds only above a threshold.
    \panel{B}~Reset-layer weights and biases from sensor-layer outputs to flexor and extensor half-centers.
    Weights start at zero, so that resetting feedback is learned.
    Columns show positive~($+$) and negative~($-$) components of each proprioceptive signal, and cell symbols show sign constraints.
}
\label{fig:a2_af_init}
\end{figure}

\paragraph{Initialization and constraints}
Sensor-layer and reset-layer biases are initialized and constrained to be nonpositive, so that each afferent responds only above a threshold.
Reset-layer weights are initialized at zero without sign constraints, so that resetting feedback is learned rather than prescribed~(\autoref{fig:a2_af_init}).
Local reset connections reflect the predominantly local action of limb afferents on spinal rhythm generators~\citep{Kim2022ContributionAfferentFeedback}.

\subsection{Pattern formation}
\label{sec:a2_pf}

\paragraph{Modules}
PF consists of forelimb~(F), hindlimb~(H), and trunk~(T) modules.
The F and H modules are shared between left and right limbs, like their AF counterparts.
The T module controls lumbar extension, bending, and twisting with 6 synergies per DOF group, receiving forelimb and hindlimb phases, brainstem command, and lumbar sensor outputs.
Bending and twisting synergies have left and right copies with sign-inverted outputs, so that mirrored inputs produce mirrored actions.

\paragraph{Delay layer}
The delay layer produces early and late copies of each limb's flexor and extensor half-center outputs with distinct delays, yielding 4 phase signals.
Delayed copies permit desynchronized joint motions within a step cycle.
Early copies use time constants~$T_v$ of 0~ms~(flexor) and 30~ms~(extensor) with bias~$B = -0.05$, and late copies use 50~ms~(flexor) and 80~ms~(extensor) with biases of 0 and $-0.3$, respectively.

\paragraph{Synergy layer}
The synergy layer recruits 5 distinct synergies: liftoff and advance during swing, and touchdown, support, and propulsion during stance.
This decomposition follows the phase-specific organization of muscle synergies in cat locomotion~\citep{Klishko2025EffectsSpinalTransection}.
The propulsion synergy has 1 copy, and the rest have 3 copies each, yielding 13 synergies per limb.
Inputs combine phase signals, local non-resetting feedback~(reflexes), and brainstem command~(speed-dependent recruitment), inspired by mutually inhibiting populations of neurons in biological pattern formation.

\paragraph{Motor layer}
The motor layer outputs a linear combination of recruited synergies to produce interface actions.
The output format depends on the neuromechanical interface~(\autoref{sec:3_neuromechanical_interfaces}).
For flexor-extensor control, each synergy specifies flexor and extensor activations for every joint of the limb.
For position control, each synergy specifies equilibrium positions for every joint of the limb.

\begin{figure}[p]
\centering
\includegraphics{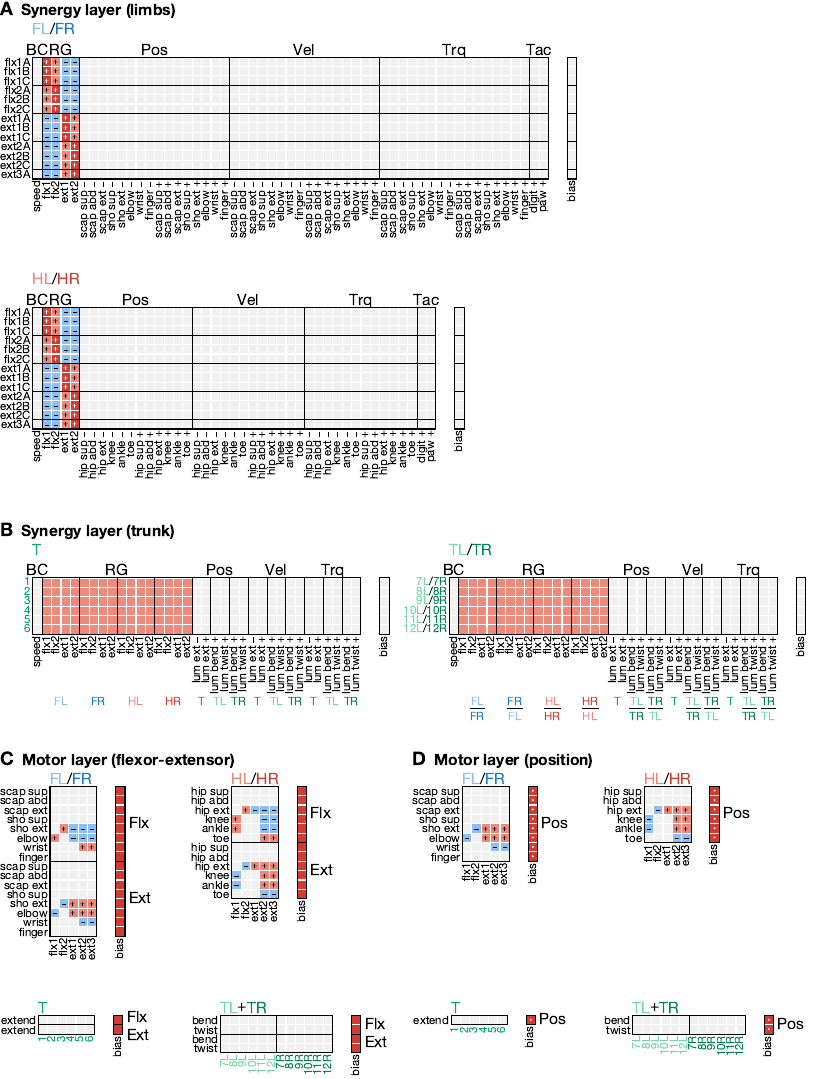}
\caption{
    \textbf{PF initialization.}
    \panel{A}~Synergy-layer parameters of the F and H modules.
    Flexor phase signals excite swing synergies and extensor phase signals excite stance synergies, each inhibiting the other group.
    \panel{B}~Synergy-layer parameters of the T module for lumbar extension and lumbar bending and twisting.
    Trunk synergies receive weak excitation from forelimb and hindlimb phases.
    \panel{C, D}~Motor-layer parameters for flexor-extensor and position control.
    The 2 interfaces share the synergy layer and differ only in the motor layer.
    Swing synergies flex and stance synergies extend the relevant joints, while biases set the equilibrium pose.
    Trunk motor-layer weights start at zero and are learned.
    Cell color shows initial values~(\textcolor{red}{positive}, \textcolor{blue}{negative}), and cell symbols show sign constraints.
}
\label{fig:a2_pf_init}
\end{figure}

\paragraph{Initialization and constraints}
Synergy-layer weights are initialized so that flexor phase signals excite swing synergies and inhibit stance synergies, and extensor phase signals do the opposite~($\pm 1$).
Motor-layer weights are sign-constrained and initialized with small values on joints consistent with each synergy's role~(\autoref{fig:a2_pf_init}).
For example, hindlimb liftoff synergies flex the knee and ankle, advance synergies flex the hip, and support and propulsion synergies extend the hip, knee, and ankle.
The copies of each synergy share an initialization and differentiate during training.
In the T module, motor-layer weights are initialized at zero and synergy-layer weights from limb phases are initialized small~(0.01).
Motor-layer biases are initialized at values that produce the interface equilibrium pose: moderate co-activation~(0.8) of flexors and extensors.
For flexor-extensor control, biases remain trainable, so that co-activation can decrease to make joints more compliant.
For position control, biases are frozen at 0.5, so that the equilibrium pose remains the default standing pose.

\clearpage
\subsection{Parameter counts}
\label{sec:a2_parameter_counts}

Rat~NCAP has 35 fixed parameters in RG and BC and 2,133--2,238 trained parameters in AF and PF~(\autoref{tab:a2_parameter_counts}).
The interfaces differ only in the motor layer, whose size depends on the number of control variables per DOF.
The MLPs have about 50 times more parameters than Rat~NCAP.

\begin{table}[h]
\centering
\caption{
    \textbf{Parameter counts.}
    Parameters of Rat~NCAP by module~(top) and of the MLP baselines by layer~(bottom) for each neuromechanical interface.
}
\label{tab:a2_parameter_counts}
\small
\setlength{\tabcolsep}{4pt}
\begin{tabular}{@{}lrrrr@{}}
\toprule
 & Torque & Position & Position-stiffness & Flexor-extensor \\
\midrule
\multicolumn{5}{@{}l}{\textbf{Rat~NCAP}} \\
RG~(fixed) & 33 & 33 & 33 & 33 \\
BC~(fixed) & 2 & 2 & 2 & 2 \\
AF sensor layer & 200 & 200 & 200 & 200 \\
AF reset layer & 180 & 180 & 180 & 180 \\
PF synergy layer: RG, BC inputs & 324 & 324 & 324 & 324 \\
PF synergy layer: AF inputs & 1,324 & 1,324 & 1,324 & 1,324 \\
PF motor layer & 105 & 105 & 210 & 193 \\
\cmidrule(l){2-5}
Trained & 2,133 & 2,133 & 2,238 & 2,221 \\
\textbf{Total} & \textbf{2,168} & \textbf{2,168} & \textbf{2,273} & \textbf{2,256} \\
\midrule
\multicolumn{5}{@{}l}{\textbf{MLP}} \\
Hidden layer 1~[102, 256] & -- & 26,368 & -- & 26,368 \\
Hidden layer 2~[256, 256] & -- & 65,792 & -- & 65,792 \\
Output layer~[256, 31 or 62] & -- & 7,967 & -- & 15,934 \\
\cmidrule(l){2-5}
\textbf{Total} & -- & \textbf{100,127} & -- & \textbf{108,094} \\
\bottomrule
\end{tabular}
\end{table}

\clearpage
\section{Neuromechanical Interfaces}
\label{sec:a3_interfaces}

\subsection{Observation transform}
\label{sec:a3_observations}

All interfaces share an observation transform that rectifies proprioceptive and tactile observations into nonnegative components, analogous to afferents that signal opposite directions of joint motion and loading.
Each joint has angle~$q$ with flexion and extension limits~$[q_\text{min}, q_\text{max}]$ and range~$R = q_\text{max} - q_\text{min}$.
Position is normalized to~$\hat{q} = (q - q_\text{min}) / R \in [0, 1]$, and its negative and positive components measure closeness to the flexion and extension limits, quantifying the amount of flexion and extension.
Velocity and torque are sign-rectified.
Tactile contact force is nonnegative and has only a positive component:
\begin{align*}
    o^{\text{pos}-} &= 1 - \hat{q}, & o^{\text{vel}-} &= \max(0, -\dot{q}), & o^{\text{trq}-} &= \max(0, -\tau), \\
    o^{\text{pos}+} &= \hat{q}, & o^{\text{vel}+} &= \max(0, \dot{q}), & o^{\text{trq}+} &= \max(0, \tau), & o^{\text{tac}+} &= f
\end{align*}

\subsection{Flexor-extensor control}
\label{sec:a3_flxext_control}

Flexor-extensor control actuates each joint with an antagonistic pair of linear springs, a flexor and an extensor, scaled by activations~$a_\text{flx}, a_\text{ext} \in [0, 1]$ output by the neural circuit.

\paragraph{Geometric interpretation}
The springs can be pictured as a pulley system acting antagonistically on the joint~(\figref[A]{fig:a3_flxext}).
Each spring wraps around a pulley of constant radius at the joint, so joint rotation changes its length linearly.
The flexor has stiffness~$k_\text{flx}$ and rests at its attachment point~$q_\text{flx} \leq q_\text{min}$, and the extensor has stiffness~$k_\text{ext}$ and rests at~$q_\text{ext} \geq q_\text{max}$.
Since the attachment points lie beyond the joint limits, both springs remain stretched throughout the range of motion and can only pull, as muscles do: the flexor always produces flexion torque, and the extensor always produces extension torque.
The muscles play a tug-of-war, with activations setting how strongly each pulls.
The joint settles where the pulls balance, or at a joint limit if the balance point lies beyond it.

\begin{figure}[h]
\centering
\includegraphics{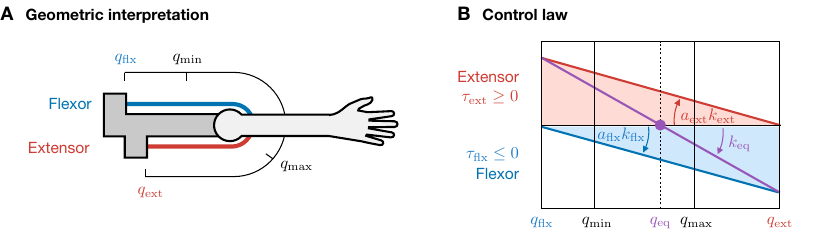}
\caption{
    \textbf{Flexor-extensor control.}
    \panel{A}~Geometric interpretation.
    \textcolor{blue}{Flexor} and \textcolor{red}{extensor} springs act on the joint like a pulley system, pulling against each other from attachment points beyond the joint limits.
    \panel{B}~Control law.
    Flexor and extensor torques versus joint angle, shaded over activations from 0 to 1.
    \textcolor{purple}{Net torque} acts as a single spring with activation-dependent stiffness and equilibrium position.
}
\label{fig:a3_flxext}
\end{figure}

\paragraph{Control law}
The joint torque is the sum of extensor and flexor torques~(\figref[B]{fig:a3_flxext}):
\begin{equation*}
    \tau = \tau_\text{ext} + \tau_\text{flx}, \qquad
    \tau_\text{ext} = a_\text{ext} \cdot k_\text{ext} \cdot (q_\text{ext} - q), \qquad
    \tau_\text{flx} = a_\text{flx} \cdot k_\text{flx} \cdot (q_\text{flx} - q)
\end{equation*}
with $\tau_\text{ext} \geq 0$ and $\tau_\text{flx} \leq 0$ within the joint limits.
Grouping terms yields position-stiffness control with activation-dependent stiffness~$k_\text{eq}$ and equilibrium position~$q_\text{eq}$:
\begin{equation*}
    \tau = k_\text{eq} \cdot (q_\text{eq} - q), \qquad
    k_\text{eq} = a_\text{ext} \cdot k_\text{ext} + a_\text{flx} \cdot k_\text{flx}, \qquad
    q_\text{eq} = \frac{a_\text{ext} \cdot k_\text{ext} \cdot q_\text{ext} + a_\text{flx} \cdot k_\text{flx} \cdot q_\text{flx}}{k_\text{eq}}
\end{equation*}
The equilibrium position is a stiffness-weighted average of the 2 attachment points.
Differential activation shifts the equilibrium position toward the more active muscle's attachment point, while co-activation scales stiffness without moving the equilibrium position.
The neural circuit can therefore control both equilibrium position and stiffness through 2 activations with fixed flexor and extensor identities, a principle reflected in biological motor control~\citep{Shadmehr1993ControlEquilibriumPosition}.
Unlike in position-stiffness control, the 2 quantities are coupled: stiffness is greatest at full co-activation and decreases as the equilibrium position approaches either attachment point.

\paragraph{Parameter matching}
Flexor-extensor control is specified by 4 spring parameters per joint~$(k_\text{flx}, k_\text{ext}, q_\text{flx}, q_\text{ext})$, whereas the other interfaces are specified by an equilibrium position, stiffness, and torque.
The spring parameters can be chosen so that flexor-extensor control matches per-joint parameters shared with the other interfaces, which enables comparing interfaces on equal footing.
These shared parameters are equilibrium position~$q_\text{eq}$, stiffness~$k_\text{eq}$, and maximum flexor and extensor torques~$\tau_\text{flx}^\text{max}, \tau_\text{ext}^\text{max}$, which serve as fixed parameters or control ranges.
Matching the 4 shared parameters provides 4 conditions, which uniquely determine the 4 spring parameters.
First, at full co-activation~($a_\text{flx} = a_\text{ext} = 1$), the net spring has the shared stiffness, and the 2 spring torques balance at the shared equilibrium position.
Second, each muscle produces its maximum torque when fully active alone at the opposite joint limit, where it is stretched furthest.
These conditions are:
\begin{align*}
    k_\text{ext} + k_\text{flx} &= k_\text{eq}
    & k_\text{ext} \cdot (q_\text{ext} - q_\text{eq}) &= k_\text{flx} \cdot (q_\text{eq} - q_\text{flx}) \\
    k_\text{ext} \cdot (q_\text{ext} - q_\text{min}) &= \tau_\text{ext}^\text{max}
    & k_\text{flx} \cdot (q_\text{max} - q_\text{flx}) &= \tau_\text{flx}^\text{max}
\end{align*}
Solving these conditions yields:
\begin{align*}
    k_\text{ext} &= \frac{k_\text{eq} \cdot (q_\text{max} - q_\text{eq}) + \tau_\text{ext}^\text{max} - \tau_\text{flx}^\text{max}}{R},
    & q_\text{ext} &= q_\text{min} + \frac{\tau_\text{ext}^\text{max}}{k_\text{ext}}, \\
    k_\text{flx} &= \frac{k_\text{eq} \cdot (q_\text{eq} - q_\text{min}) + \tau_\text{flx}^\text{max} - \tau_\text{ext}^\text{max}}{R},
    & q_\text{flx} &= q_\text{max} - \frac{\tau_\text{flx}^\text{max}}{k_\text{flx}}.
\end{align*}
The derived parameters are valid when $k_\text{ext}, k_\text{flx} > 0$ and the attachment points lie beyond the joint limits, which requires $\tau_\text{ext}^\text{max} \geq k_\text{ext} R$ and $\tau_\text{flx}^\text{max} \geq k_\text{flx} R$.
That is, each maximum torque must be large enough relative to its stiffness that the spring remains stretched, and thus pulls, across the full range of motion.
In this work, $q_\text{eq}$ is the midpoint of the joint range and $\tau_\text{ext}^\text{max} = \tau_\text{flx}^\text{max}$, so that $k_\text{ext} = k_\text{flx} = k_\text{eq}/2$ and the attachment points are placed symmetrically about the range.
Per-joint values of the shared parameters are provided in the released configuration files.

\paragraph{Comparison to Ekeberg model}
The most closely related abstract muscle model is that of \citet{Ekeberg1993CombinedNeuronalMechanical}, as formulated in \citet{Lobato-Rios2022NeuroMechFlyNeuromechanicalModel}.
Up to sign conventions and omitting passive and damping terms, it sums a differential term with gain~$k_a$ and a co-activation-dependent stiffness term with gain~$k_s$ around a rest position~$q_\text{rest}$.
Grouping terms by activation yields 2 linear springs:
\begin{align*}
    \tau &= k_a \cdot (a_\text{ext} - a_\text{flx}) + k_s \cdot (a_\text{ext} + a_\text{flx}) \cdot (q_\text{rest} - q) \\
    &= a_\text{ext} \cdot k_s \cdot \bigl((q_\text{rest} + k_a / k_s) - q\bigr) + a_\text{flx} \cdot k_s \cdot \bigl((q_\text{rest} - k_a / k_s) - q\bigr)
\end{align*}
This is the flexor-extensor control law with symmetric stiffnesses~$k_\text{flx} = k_\text{ext} = k_s$ and attachment points~$q_\text{ext}, q_\text{flx} = q_\text{rest} \pm k_a / k_s$.
The difference is that $k_a$ and $k_s$ are independent in the Ekeberg model, so its attachment points are not constrained to lie beyond the joint limits.
When an attachment point lies within the range of motion, the muscle reverses its torque past that point.
For example, if $q_\text{flx} > q_\text{min}$, flexor activation produces extension torque at angles between $q_\text{min}$ and $q_\text{flx}$, so the circuit must account for a state-dependent sign of each muscle's effect.
Our flexor-extensor model rules out this sign reversal by construction and permits asymmetric flexor and extensor strengths.

\paragraph{Extensions}
The flexor-extensor interface is a starting point for more realistic muscle models.
Since each joint has explicit flexor and extensor activations, Hill-type muscle properties, such as activation dynamics, force-length curves, and force-velocity relationships, can be added one at a time.
Their contributions to circuit function and behavior can then be tested against the linear springs used here.

\clearpage
\section{Training Setup}
\label{sec:a4_training}

\subsection{Tasks}
\label{sec:a4_tasks}

\paragraph{Initialization}
Each episode begins with the rat body in the default standing pose at the origin, with zero velocity.
The control timestep is $\Delta t_\text{env} = 0.03$~s, and the physics timestep is $\Delta t_\text{sim} = 0.001$~s~(30 substeps per control step).
Training episodes are truncated after 100 control steps~(3~s), and test episodes after 200 control steps~(6~s).

\paragraph{Termination}
An episode terminates early if the trunk or upper limbs touch the ground, or if roll or pitch exceeds $45^\circ$.
These conditions detect falls and prevent prone or rolling behaviors.

\paragraph{Subtasks}
The locomotion task has 6 subtasks, each with the goal of matching a fixed target speed~\mbox{($v^\text{target} \in \{0.0, 0.2, 0.4, 0.6, 0.8, 1.0\}$~m/s)}.
Each population member is evaluated on all 6 subtasks in parallel, and its fitness aggregates returns across subtasks.
The target speed is provided to the model at each timestep through the brainstem command $c_t = v^\text{target}_t$.

\paragraph{Rewards}
For moving subtasks~($v^\text{target} > 0$), the reward is a weighted sum of 3 terms:
\begin{equation*}
    r_t = w_\text{alive}\, r_\text{alive} + w_\text{dense}\, r_\text{dense}(v_t, v^\text{target}_t) + w_\text{sparse}\, r_\text{sparse}(v_t, v^\text{target}_t)
\end{equation*}
with forward velocity~$v_t$ and weights $w_\text{alive} = 0.2$, $w_\text{dense} = 0.6$, and $w_\text{sparse} = 0.2$.
For the standing subtask~($v^\text{target} = 0$), the reward is a pose term~$r_t = r_\text{pose}$.
All terms lie in $[0, 1]$.

An \textbf{alive reward}~$r_\text{alive} = 1$ is given at every non-terminated timestep, which discourages early termination without shaping behavior.

A \textbf{dense velocity reward} is a triangular function centered at the target speed:
\begin{equation*}
    r_\text{dense}(v_t, v^\text{target}_t) = \max\left(0,\, 1 - \frac{|v_t - v^\text{target}_t|}{\max(v^\text{target}_t,\, \epsilon)}\right)
\end{equation*}
with margin~$\epsilon = 0.1$~m/s.
It reaches 1 when $v_t = v^\text{target}_t$ and declines linearly to 0 at a deviation of $\max(v^\text{target}_t, \epsilon)$, providing a graded signal toward the target speed.

A \textbf{sparse velocity reward} is a quadratic function centered at the target speed:
\begin{equation*}
    r_\text{sparse}(v_t, v^\text{target}_t) = \max\left(0,\, 1 - \left(\frac{|v_t - v^\text{target}_t|}{\epsilon}\right)^2\right)
\end{equation*}
It reaches 1 when $v_t = v^\text{target}_t$ and declines quadratically to 0 at a deviation of $\epsilon$, rewarding consistency near the target speed.

A \textbf{pose reward} decays with the largest joint-angle deviation from the default standing pose:
\begin{equation*}
    r_\text{pose} = \frac{1}{1 + 4\,(d_t / \epsilon_\text{pose})^2}, \qquad d_t = \max_j |q_{t,j} - q^\text{stand}_j|
\end{equation*}
with joint angles~$q_{t,j}$, standing-pose joint angles~$q^\text{stand}_j$, and margin~$\epsilon_\text{pose} = 10^\circ$.

\subsection{Training algorithm}
\label{sec:a4_algorithm}

We optimize the trainable AF and PF parameters~$\vtheta$ with Augmented Random Search~(ARS)~\citep{Mania2018SimpleRandomSearch}, implemented on GPU alongside the vectorized environments.
Evolutionary strategies such as ARS suit biologically grounded architectures because they require neither backpropagation through recurrent neural dynamics and simulator contacts nor a differentiable model, and they handle sign constraints and frozen parameters directly.
ARS in particular has few hyperparameters and is competitive with deep reinforcement learning on continuous control benchmarks~\citep{Mania2018SimpleRandomSearch}.

The return of parameters~$\vtheta$ aggregates rewards over all subtasks and timesteps of an episode:
\begin{equation*}
    R(\vtheta) = \frac{1}{|\mathcal{V}|} \sum_{v^\text{target} \in \mathcal{V}} \frac{1}{T} \sum_{t=1}^{T} r_t
\end{equation*}
with set of target speeds~$\mathcal{V}$, episode length~$T$, and rewards~$r_t$ obtained by the model with parameters~$\vtheta$.
Each epoch samples random directions~$\vdelta_i \sim \mathcal{N}(\mathbf{0}, \mI)$, evaluates the perturbed parameters~$\vtheta \pm \nu \vdelta_i$, and updates parameters along the $b$ best-performing directions:
\begin{equation*}
    \vtheta \leftarrow \vtheta + \frac{\alpha}{b \, \sigma_R} \sum_{i \in \text{top-}b} \big( R(\vtheta + \nu \vdelta_i) - R(\vtheta - \nu \vdelta_i) \big) \, \vdelta_i
\end{equation*}
with perturbation scale~$\nu$, step size~$\alpha$, and standard deviation~$\sigma_R$ of the returns used in the update.
Sign and range constraints are re-applied after each update.
For hyperparameters, see~\autoref{tab:a4_ars}.

\begin{table}[h]
\centering
\caption{\textbf{ARS hyperparameters.}}
\label{tab:a4_ars}
\small
\begin{tabular}{@{}lr@{}}
\toprule
Hyperparameter & Value \\
\midrule
Population size & 1024 \\
Top directions~$b$ & 256 \\
Perturbation scale~$\nu$ & 0.02 \\
Step size~$\alpha$ & 0.03 \\
Epochs & 4000 \\
Training rollout length & 100 steps~(3~s) \\
Test rollout length & 200 steps~(6~s) \\
Subtasks per population member & 6 \\
\bottomrule
\end{tabular}
\end{table}

\subsection{Computational resources}
\label{sec:a4_computational_resources}

All experiments were run on a Linux cluster with 4 CPU cores, 16~GB of RAM, and 1 GPU~(NVIDIA L40S, 48~GB VRAM) per run.
Each training run takes about 36~h.
The reported results comprise 19 training runs: Rat~NCAP with 3 seeds for each of the 4 interfaces, 2 DOF ablations of Rat~NCAP with 1 seed each, the ES-trained and PPO-trained MLPs with 1 seed each for the position and flexor-extensor interfaces, and Quadruped~NCAP with 1 seed.
Training totals about 684 GPU-hours, and bumpy-terrain tests and circuit ablations add about 4 GPU-hours.

\clearpage
\section{Analysis Methods}
\label{sec:a5_analysis}

\subsection{Behavioral measures}
\label{sec:a5_behavioral_measures}

\paragraph{Evaluation protocol}
Behavioral measures are computed for 3 seeds from 15-s evaluation rollouts, using the final 10~s after a 5-s settling period~(\autoref{fig:5_experimental_validation} and~\autoref{fig:7_cell_type_ablations}), and from the 6-s test episodes~(\autoref{fig:6_neuromechanical_comparison}).

\paragraph{Step statistics}
Stance is defined as paw touch-sensor contact associated with the limb's RG extension phase, which removes brief contacts during swing and fills brief lift-offs during stance.
A step cycle spans consecutive touchdowns of the same paw.
Stance and swing durations are measured within each cycle, step frequency is the inverse of cycle duration, and step length is the horizontal distance traveled by the paw during swing~(\figref[A--C]{fig:5_experimental_validation}).
Values outside 1.5 times the interquartile range at each target speed are excluded as outliers.

\paragraph{Gait classification}
Interlimb phases are computed for each step cycle as the touchdown time of each limb relative to the right forelimb, normalized by cycle duration.
Each cycle is classified by comparing its interlimb phases with idealized phases: diagonal synchrony with left-right alternation for trot, left-right synchrony for bound, and sequential quarter-cycle offsets for walk, within a tolerance of 0.25 cycles for trot and bound and 0.125 cycles for walk.
Gait classification reports the fraction of step cycles matching each gait~(\autoref{fig:7_cell_type_ablations}).

\paragraph{Kinetics and kinematics}
Ground reaction forces are computed from paw contact forces and normalized by body mass~(\figref[F]{fig:5_experimental_validation}).
Ground reaction forces and joint angles~(\figref[G]{fig:5_experimental_validation}) are resampled to 100 points over a normalized step cycle from touchdown to touchdown, then averaged across cycles.

\paragraph{Mechanical power}
The mechanical power of actuator~$j$ at timestep~$t$ is $p_{j,t} = \tau_{j,t}\, \dot{q}_{j,t}$, with actuator force~$\tau_{j,t}$ and actuator velocity~$\dot{q}_{j,t}$.
Positive and negative mechanical power are the sums of $\max(p_{j,t}, 0)$ and $\min(p_{j,t}, 0)$ over limb actuators, averaged over timesteps~(\figref[C, D]{fig:6_neuromechanical_comparison}).
Positive power reflects propulsive work, and negative power reflects energy absorbed by the joints, such as during braking.

\subsection{Circuit perturbations}
\label{sec:a5_circuit_perturbations}

\paragraph{Evaluation protocol}
Perturbations are applied to trained models at test time without retraining, at 0.4~m/s, where the intact model trots, with 30 trials for each of 3 seeds.

\paragraph{RG weight noise}
Each RG synaptic weight~$w$ is perturbed multiplicatively as $(1 + \sigma \epsilon)\, w$ with $\epsilon \sim \mathcal{N}(0, 1)$, clipped so that each weight preserves its sign~(\figref[A]{fig:7_cell_type_ablations}).
Noise is sampled independently for each trial, and $\sigma$ ranges from 0 to 0.5 in increments of 0.025.

\paragraph{Cell-type activity scaling}
The outputs of targeted RG populations are multiplied by an activity scale~$k$ before projecting to their targets, with $k = 1$ for the intact circuit and $k = 0$ for complete ablation.
Scaling is applied jointly to V0v and V0d populations~(\figref[B]{fig:7_cell_type_ablations}) or to V3f and V3a populations~(\figref[C]{fig:7_cell_type_ablations}), with $k$ ranging from 1 to 0 in increments of 0.05.

\clearpage
\section{Architecture Baselines}
\label{sec:a6_baselines}

Generic architectures serve as scientific controls rather than benchmarks, since a generic controller with higher return would still not map onto experimentally identified cell types or connectivity~\citep{Cao2024ExplanatoryModelsNeuroscience}.
MLPs test whether the interface or training algorithm alone explains Rat~NCAP's behavior, and Quadruped~NCAP tests whether its architectural additions are critical.

\subsection{MLP training}
\label{sec:a6_mlp_training}

We train 2 MLPs with the same observations, interfaces, rewards, and target speeds as Rat~NCAP.
The ES-trained MLP uses the same ARS algorithm and hyperparameters as Rat~NCAP, isolating the effect of architecture.
The PPO-trained MLP uses Proximal Policy Optimization~(PPO)~\citep{Schulman2017ProximalPolicyOptimization}, a standard deep reinforcement learning algorithm, providing a reference for reasonable return.
The ES-trained MLP uses 2 hidden layers of 256 units with ReLU activations and outputs clipped to the action range.
The PPO-trained MLP uses an actor with the same architecture and a Gaussian policy with a learned, state-independent standard deviation, and a separate critic with 2 hidden layers of 256 units with tanh activations.
Both MLPs receive the target speed concatenated with proprioceptive and tactile observations.
For PPO hyperparameters, see~\autoref{tab:a6_ppo}.

\begin{table}[h]
\centering
\caption{\textbf{PPO hyperparameters.}}
\label{tab:a6_ppo}
\small
\begin{tabular}{@{}lr@{\hspace{2em}}lr@{}}
\toprule
Hyperparameter & Value & Hyperparameter & Value \\
\midrule
Parallel environments & 6144 & KL early-stop threshold & 0.05 \\
Steps per update & 40 & Optimizer & Adam \\
Discount & 0.97 & Learning rate & $3 \times 10^{-4}$ \\
GAE~$\lambda$ & 0.95 & Entropy coefficient & 0 \\
Minibatch size & 4096 & Gradient clipping & None \\
Minibatch iterations & 4 & Advantage normalization & Yes \\
Clip ratio & 0.2 & &  \\
\bottomrule
\end{tabular}
\end{table}

\begin{figure}[h]
\centering
\includegraphics{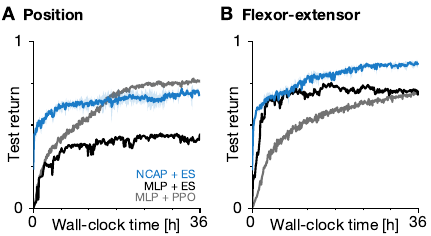}
\caption{
    \textbf{MLP training.}
    Test return over wall-clock training time for Rat~NCAP and ES- or PPO-trained MLPs.
    \panel{A}~Position control.
    The PPO-trained MLP eventually slightly exceeds Rat~NCAP, while the ES-trained MLP remains well below.
    \panel{B}~Flexor-extensor control.
    Rat~NCAP reaches the highest return.
    Rat~NCAP data over 3 seeds with mean~$\pm$~SD; MLP data from 1 seed each.
}
\label{fig:a6_mlp_training}
\end{figure}

Rat~NCAP achieves return comparable to or higher than both MLPs~(\autoref{fig:a6_mlp_training}).
With flexor-extensor control, Rat~NCAP reaches the highest return, while the ES-trained MLP plateaus lower and the PPO-trained MLP improves more slowly to a similar final return.
With position control, the PPO-trained MLP eventually reaches slightly higher return than Rat~NCAP, while the ES-trained MLP plateaus substantially lower.
With both interfaces, Rat~NCAP reaches high return early in training, consistent with the inductive bias of its architecture.
Since the same ARS algorithm yields lower return for an MLP, the performance of Rat~NCAP does not arise from the training algorithm alone.

\clearpage
\subsection{MLP behavior}
\label{sec:a6_mlp_behavior}

\begin{figure}[h]
\centering
\includegraphics{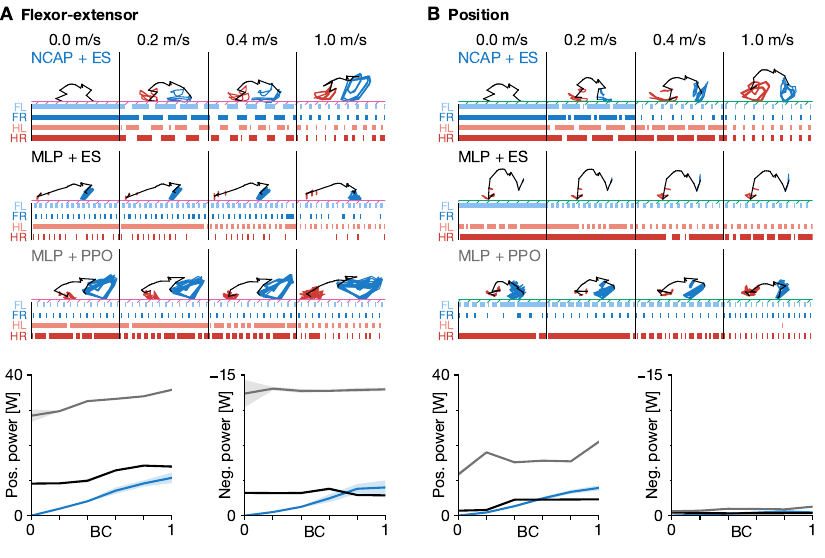}
\caption{
    \textbf{MLP behavior.}
    Limb trajectories and footfall patterns at representative target speeds~(top) and mean positive and negative mechanical power as brainstem command increases~(bottom) for Rat~NCAP~(\textcolor{blue}{blue}) and MLPs trained with ES~(black) or PPO~(\textcolor{gray}{gray}).
    \panel{A}~Flexor-extensor control.
    \panel{B}~Position control.
    Rat~NCAP stands at 0~m/s and shifts gaits with speed, while the MLPs step irregularly at all speeds and the PPO-trained MLP expends far more mechanical power.
}
\label{fig:a6_mlp_behavior}
\end{figure}

Despite competitive return in some conditions, both MLPs produce markedly less naturalistic behavior than Rat~NCAP with both interfaces~(\autoref{fig:a6_mlp_behavior}).
Rat~NCAP stands still at 0~m/s and transitions from walk to trot to bound as target speed increases, with mechanical power rising with speed.
The MLPs instead step continuously even at 0~m/s, with irregular footfall patterns and no distinct gait transitions.
The ES-trained MLP adopts abnormal postures with small, stereotyped steps, while the PPO-trained MLP produces large, erratic limb movements and expends substantially more mechanical power than Rat~NCAP at every target speed.
These results are consistent with prior work in this rat, in which generic architectures trained without imitation data produced unrealistic movement~\citep{Merel2019DeepNeuroethologyVirtual}.
Together, these controls indicate that neither the neuromechanical interface nor the training algorithm alone explains the naturalistic behavior of Rat~NCAP.

\clearpage
\subsection{MLP generalization}
\label{sec:a6_mlp_generalization}

\begin{figure}[h]
\centering
\includegraphics{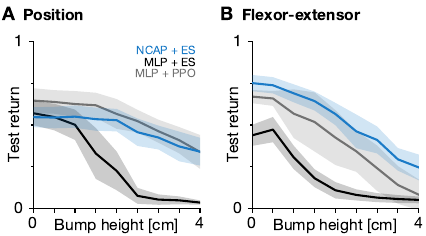}
\caption{
    \textbf{MLP generalization.}
    Test return of models trained on flat ground and evaluated on terrain with increasing bump height.
    \panel{A}~Position control.
    Rat~NCAP performs comparably to the PPO-trained MLP, while the ES-trained MLP collapses.
    \panel{B}~Flexor-extensor control.
    Rat~NCAP retains the highest return at all bump heights.
    Mean~$\pm$~SD over 128 episodes per bump height.
}
\label{fig:a6_mlp_generalization}
\end{figure}

Rat~NCAP generalizes to unseen bumpy terrain at least as well as both MLPs with both interfaces~(\autoref{fig:a6_mlp_generalization}).
All models are trained only on flat ground and evaluated on terrain with randomly placed bumps of increasing height.
The terrain is a heightfield constructed by drawing uniform random heights on a 5-cm grid and upsampling them with cubic-spline interpolation to a 2-cm mesh, producing smooth bumps about 5~cm across.
Heights are scaled so that bump peaks lie at ground level and troughs lie one bump height below, with bump heights from 0 to 4~cm in 0.5-cm increments.
With position control, Rat~NCAP performs comparably to the PPO-trained MLP, while the ES-trained MLP collapses on bumps above about 1~cm.
With flexor-extensor control, Rat~NCAP achieves the highest return at all bump heights, while the ES-trained MLP drops sharply even on small bumps.

This experiment evaluates transfer to bumpy terrain, not general robustness to perturbations, obstacles, or sensory conditions.
Such settings would likely also require vestibular and visual feedback and additional circuits such as the cerebellum, which are outside the scope of the present spinal model.

\subsection{Quadruped NCAP training}
\label{sec:a6_quadruped_training}

\begin{figure}[h]
\centering
\includegraphics{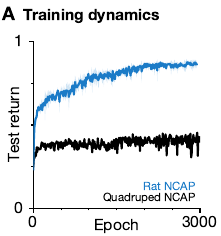}
\caption{
    \textbf{Quadruped NCAP training.}
    Test return over training epochs for Rat~NCAP and Quadruped~NCAP on Rat~v2.
    Quadruped~NCAP plateaus early at about half the return of Rat~NCAP.
    Rat~NCAP data over 3 seeds with mean~$\pm$~SD; Quadruped~NCAP data from 1 seed.
}
\label{fig:a6_quadruped_baseline}
\end{figure}

Rat~NCAP achieves about twice the return of Quadruped~NCAP on Rat~v2~(\autoref{fig:a6_quadruped_baseline}).
Quadruped~NCAP~\citep{Bhattasali2024NeuralCircuitArchitectural} is trained with the same rewards and ARS algorithm, and its AF and PF modules are adapted to the rat joints using the same sign-constraint priors as Rat~NCAP.
It plateaus early in training.
This gap indicates that the architectural additions of Rat~NCAP, including the 2-layer AF module, phase-specific synergies, and trunk control, are collectively critical for scaling biologically grounded architectures from a simplified quadruped to a biorealistic rat body.

\clearpage
\section{Body Ablations}
\label{sec:a7_ablations}

Rat~v2 has non-planar limb DOFs and lumbar DOFs that Quadruped~NCAP did not control.
Quadruped~NCAP controlled only the planar hip and knee DOFs of each leg, since planar control suffices for locomotion in a quadruped robot and actuating hip abduction was neither necessary nor desirable~\citep{Bhattasali2024NeuralCircuitArchitectural}.
Ablating each DOF group tests its contribution to locomotion.

\subsection{DOF ablations}
\label{sec:a7_dof_ablations}

We ablate 2 DOF groups: non-planar limb DOFs~(scapula supination and abduction, shoulder supination, and hip supination and abduction) and 3 lumbar DOFs~(extension, bending, and twisting).
Each group is held near the default pose by passive spring-damper torques, as for the head, cervical, and tail DOFs, and Rat~NCAP is retrained.

Constraining non-planar limb DOFs substantially lowers test return~(\figref[B]{fig:a7_dof_ablation}).
Unlike in the quadruped robot, these DOFs are therefore important for locomotion in a biorealistic body.
In contrast, constraining lumbar DOFs does not lower test return~(\figref[A]{fig:a7_dof_ablation}).

\begin{figure}[h]
\centering
\includegraphics{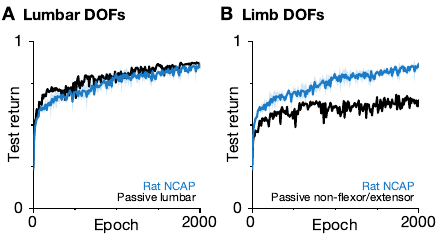}
\caption{
    \textbf{DOF ablations.}
    Test return over training epochs for Rat~NCAP with all DOFs active and with a DOF group held by passive springs.
    \panel{A}~Lumbar DOFs.
    Return is unaffected, suggesting compensation through other DOFs.
    \panel{B}~Non-planar limb DOFs.
    Return drops substantially, indicating that these DOFs are important for locomotion.
    Rat~NCAP data over 3 seeds with mean~$\pm$~SD; DOF ablation data from 1 seed each.
}
\label{fig:a7_dof_ablation}
\end{figure}

\clearpage
\subsection{Lumbar kinematics}
\label{sec:a7_lumbar_kinematics}

Although constraining lumbar DOFs does not lower return, Rat~NCAP uses them in a speed-dependent manner when available~(\figref[A]{fig:a7_lumbar_kinematics}).
At low speeds, lumbar bending and twisting oscillate once per step cycle through alternating flexor and extensor activations.
As speed increases, bending and twisting diminish and vanish in bound, while lumbar extension oscillates with larger amplitude.
With passive lumbar DOFs, these oscillations are largely absent~(\figref[B]{fig:a7_lumbar_kinematics}), so Rat~NCAP likely compensates through other DOFs.

Thus, some DOFs admit compensation while others do not, and test return alone may not reveal how the body is used.
Comparing these lumbar kinematics with experimental measurements of trunk motion in rats is an important direction for future work.

\begin{figure}[h]
\centering
\includegraphics{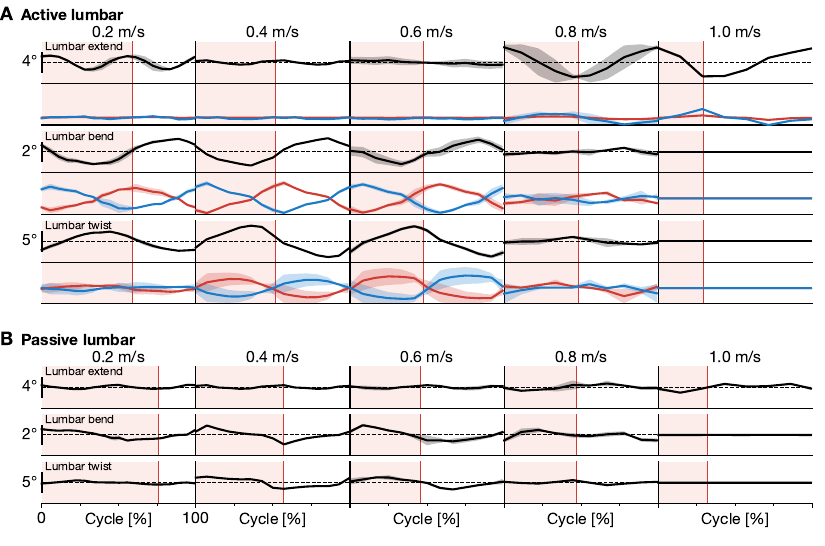}
\caption{
    \textbf{Lumbar kinematics.}
    Lumbar joint angles~(black) and \textcolor{blue}{flexor} and \textcolor{red}{extensor} activations~(\textbf{A} only) over the hindlimb step cycle at each nonzero target speed, with stance shaded and liftoff marked.
    \panel{A}~Active lumbar DOFs.
    Bending and twisting dominate at low speeds, while extension dominates at high speeds.
    \panel{B}~Passive lumbar DOFs.
    Lumbar motion remains near the default pose throughout the step cycle.
    Data from 1 seed per panel with mean~$\pm$~SD over step cycles.
}
\label{fig:a7_lumbar_kinematics}
\end{figure}

\clearpage
\section*{NeurIPS Paper Checklist}

\begin{enumerate}[leftmargin=0.25in, nosep, parsep=1ex, itemsep=0ex]

\item {\bf Claims}
    \item[] Question: Do the main claims made in the abstract and introduction accurately reflect the paper's contributions and scope?
    \item[] Answer: \answerYes{}
    \item[] Justification: The abstract and introduction enumerate contributions and the main empirical results.
    \item[] Guidelines:
    \begin{itemize}
        \item The answer \answerNA{} means that the abstract and introduction do not include the claims made in the paper.
        \item The abstract and/or introduction should clearly state the claims made, including the contributions made in the paper and important assumptions and limitations. A \answerNo{} or \answerNA{} answer to this question will not be perceived well by the reviewers. 
        \item The claims made should match theoretical and experimental results, and reflect how much the results can be expected to generalize to other settings. 
        \item It is fine to include aspirational goals as motivation as long as it is clear that these goals are not attained by the paper. 
    \end{itemize}

\item {\bf Limitations}
    \item[] Question: Does the paper discuss the limitations of the work performed by the authors?
    \item[] Answer: \answerYes{}
    \item[] Justification: Limitations are discussed in the Discussion~(\autoref{sec:5_discussion}).
    \item[] Guidelines:
    \begin{itemize}
        \item The answer \answerNA{} means that the paper has no limitation while the answer \answerNo{} means that the paper has limitations, but those are not discussed in the paper. 
        \item The authors are encouraged to create a separate ``Limitations'' section in their paper.
        \item The paper should point out any strong assumptions and how robust the results are to violations of these assumptions (e.g., independence assumptions, noiseless settings, model well-specification, asymptotic approximations only holding locally). The authors should reflect on how these assumptions might be violated in practice and what the implications would be.
        \item The authors should reflect on the scope of the claims made, e.g., if the approach was only tested on a few datasets or with a few runs. In general, empirical results often depend on implicit assumptions, which should be articulated.
        \item The authors should reflect on the factors that influence the performance of the approach. For example, a facial recognition algorithm may perform poorly when image resolution is low or images are taken in low lighting. Or a speech-to-text system might not be used reliably to provide closed captions for online lectures because it fails to handle technical jargon.
        \item The authors should discuss the computational efficiency of the proposed algorithms and how they scale with dataset size.
        \item If applicable, the authors should discuss possible limitations of their approach to address problems of privacy and fairness.
        \item While the authors might fear that complete honesty about limitations might be used by reviewers as grounds for rejection, a worse outcome might be that reviewers discover limitations that aren't acknowledged in the paper. The authors should use their best judgment and recognize that individual actions in favor of transparency play an important role in developing norms that preserve the integrity of the community. Reviewers will be specifically instructed to not penalize honesty concerning limitations.
    \end{itemize}

\item {\bf Theory assumptions and proofs}
    \item[] Question: For each theoretical result, does the paper provide the full set of assumptions and a complete (and correct) proof?
    \item[] Answer: \answerNA{}
    \item[] Justification: The paper does not include theoretical results; all contributions are empirical.
    \item[] Guidelines:
    \begin{itemize}
        \item The answer \answerNA{} means that the paper does not include theoretical results. 
        \item All the theorems, formulas, and proofs in the paper should be numbered and cross-referenced.
        \item All assumptions should be clearly stated or referenced in the statement of any theorems.
        \item The proofs can either appear in the main paper or the supplemental material, but if they appear in the supplemental material, the authors are encouraged to provide a short proof sketch to provide intuition. 
        \item Inversely, any informal proof provided in the core of the paper should be complemented by formal proofs provided in appendix or supplemental material.
        \item Theorems and Lemmas that the proof relies upon should be properly referenced. 
    \end{itemize}

    \item {\bf Experimental result reproducibility}
    \item[] Question: Does the paper fully disclose all the information needed to reproduce the main experimental results of the paper to the extent that it affects the main claims and/or conclusions of the paper (regardless of whether the code and data are provided or not)?
    \item[] Answer: \answerYes{}
    \item[] Justification: The neural circuit architecture, neuromechanical interfaces, training algorithm, reward functions, and hyperparameters are described in the Methods~(\autoref{sec:3_methods}) and Appendices~(\autoref{sec:a1_biomechanical}--\autoref{sec:a7_ablations}). Code is released at \website.
    \item[] Guidelines:
    \begin{itemize}
        \item The answer \answerNA{} means that the paper does not include experiments.
        \item If the paper includes experiments, a \answerNo{} answer to this question will not be perceived well by the reviewers: Making the paper reproducible is important, regardless of whether the code and data are provided or not.
        \item If the contribution is a dataset and\slash or model, the authors should describe the steps taken to make their results reproducible or verifiable. 
        \item Depending on the contribution, reproducibility can be accomplished in various ways. For example, if the contribution is a novel architecture, describing the architecture fully might suffice, or if the contribution is a specific model and empirical evaluation, it may be necessary to either make it possible for others to replicate the model with the same dataset, or provide access to the model. In general. releasing code and data is often one good way to accomplish this, but reproducibility can also be provided via detailed instructions for how to replicate the results, access to a hosted model (e.g., in the case of a large language model), releasing of a model checkpoint, or other means that are appropriate to the research performed.
        \item While NeurIPS does not require releasing code, the conference does require all submissions to provide some reasonable avenue for reproducibility, which may depend on the nature of the contribution. For example
        \begin{enumerate}
            \item If the contribution is primarily a new algorithm, the paper should make it clear how to reproduce that algorithm.
            \item If the contribution is primarily a new model architecture, the paper should describe the architecture clearly and fully.
            \item If the contribution is a new model (e.g., a large language model), then there should either be a way to access this model for reproducing the results or a way to reproduce the model (e.g., with an open-source dataset or instructions for how to construct the dataset).
            \item We recognize that reproducibility may be tricky in some cases, in which case authors are welcome to describe the particular way they provide for reproducibility. In the case of closed-source models, it may be that access to the model is limited in some way (e.g., to registered users), but it should be possible for other researchers to have some path to reproducing or verifying the results.
        \end{enumerate}
    \end{itemize}

\item {\bf Open access to data and code}
    \item[] Question: Does the paper provide open access to the data and code, with sufficient instructions to faithfully reproduce the main experimental results, as described in supplemental material?
    \item[] Answer: \answerYes{}
    \item[] Justification: Code is released at \website.
    \item[] Guidelines:
    \begin{itemize}
        \item The answer \answerNA{} means that paper does not include experiments requiring code.
        \item Please see the NeurIPS code and data submission guidelines (\url{https://neurips.cc/public/guides/CodeSubmissionPolicy}) for more details.
        \item While we encourage the release of code and data, we understand that this might not be possible, so \answerNo{} is an acceptable answer. Papers cannot be rejected simply for not including code, unless this is central to the contribution (e.g., for a new open-source benchmark).
        \item The instructions should contain the exact command and environment needed to run to reproduce the results. See the NeurIPS code and data submission guidelines (\url{https://neurips.cc/public/guides/CodeSubmissionPolicy}) for more details.
        \item The authors should provide instructions on data access and preparation, including how to access the raw data, preprocessed data, intermediate data, and generated data, etc.
        \item The authors should provide scripts to reproduce all experimental results for the new proposed method and baselines. If only a subset of experiments are reproducible, they should state which ones are omitted from the script and why.
        \item At submission time, to preserve anonymity, the authors should release anonymized versions (if applicable).
        \item Providing as much information as possible in supplemental material (appended to the paper) is recommended, but including URLs to data and code is permitted.
    \end{itemize}

\item {\bf Experimental setting/details}
    \item[] Question: Does the paper specify all the training and test details (e.g., data splits, hyperparameters, how they were chosen, type of optimizer) necessary to understand the results?
    \item[] Answer: \answerYes{}
    \item[] Justification: The details and hyperparameters of the model, algorithm, and environment are described in the Appendices.
    \item[] Guidelines:
    \begin{itemize}
        \item The answer \answerNA{} means that the paper does not include experiments.
        \item The experimental setting should be presented in the core of the paper to a level of detail that is necessary to appreciate the results and make sense of them.
        \item The full details can be provided either with the code, in appendix, or as supplemental material.
    \end{itemize}

\item {\bf Experiment statistical significance}
    \item[] Question: Does the paper report error bars suitably and correctly defined or other appropriate information about the statistical significance of the experiments?
    \item[] Answer: \answerYes{}
    \item[] Justification: Training curves and quantitative comparisons report error bars and figure captions state the source of variability.
    \item[] Guidelines:
    \begin{itemize}
        \item The answer \answerNA{} means that the paper does not include experiments.
        \item The authors should answer \answerYes{} if the results are accompanied by error bars, confidence intervals, or statistical significance tests, at least for the experiments that support the main claims of the paper.
        \item The factors of variability that the error bars are capturing should be clearly stated (for example, train/test split, initialization, random drawing of some parameter, or overall run with given experimental conditions).
        \item The method for calculating the error bars should be explained (closed form formula, call to a library function, bootstrap, etc.)
        \item The assumptions made should be given (e.g., Normally distributed errors).
        \item It should be clear whether the error bar is the standard deviation or the standard error of the mean.
        \item It is OK to report 1-sigma error bars, but one should state it. The authors should preferably report a 2-sigma error bar than state that they have a 96\% CI, if the hypothesis of Normality of errors is not verified.
        \item For asymmetric distributions, the authors should be careful not to show in tables or figures symmetric error bars that would yield results that are out of range (e.g., negative error rates).
        \item If error bars are reported in tables or plots, the authors should explain in the text how they were calculated and reference the corresponding figures or tables in the text.
    \end{itemize}

\item {\bf Experiments compute resources}
    \item[] Question: For each experiment, does the paper provide sufficient information on the computer resources (type of compute workers, memory, time of execution) needed to reproduce the experiments?
    \item[] Answer: \answerYes{}
    \item[] Justification: Computational resources, including per-run training time, are reported in~\autoref{sec:a4_computational_resources}.
    \item[] Guidelines:
    \begin{itemize}
        \item The answer \answerNA{} means that the paper does not include experiments.
        \item The paper should indicate the type of compute workers CPU or GPU, internal cluster, or cloud provider, including relevant memory and storage.
        \item The paper should provide the amount of compute required for each of the individual experimental runs as well as estimate the total compute. 
        \item The paper should disclose whether the full research project required more compute than the experiments reported in the paper (e.g., preliminary or failed experiments that didn't make it into the paper). 
    \end{itemize}
    
\item {\bf Code of ethics}
    \item[] Question: Does the research conducted in the paper conform, in every respect, with the NeurIPS Code of Ethics \url{https://neurips.cc/public/EthicsGuidelines}?
    \item[] Answer: \answerYes{}
    \item[] Justification: This work involves computational simulation only and raises no ethical concerns.
    \item[] Guidelines:
    \begin{itemize}
        \item The answer \answerNA{} means that the authors have not reviewed the NeurIPS Code of Ethics.
        \item If the authors answer \answerNo, they should explain the special circumstances that require a deviation from the Code of Ethics.
        \item The authors should make sure to preserve anonymity (e.g., if there is a special consideration due to laws or regulations in their jurisdiction).
    \end{itemize}

\item {\bf Broader impacts}
    \item[] Question: Does the paper discuss both potential positive societal impacts and negative societal impacts of the work performed?
    \item[] Answer: \answerNA{}
    \item[] Justification: This is foundational neuroscience research with no foreseeable direct path to harmful applications.
    \item[] Guidelines:
    \begin{itemize}
        \item The answer \answerNA{} means that there is no societal impact of the work performed.
        \item If the authors answer \answerNA{} or \answerNo, they should explain why their work has no societal impact or why the paper does not address societal impact.
        \item Examples of negative societal impacts include potential malicious or unintended uses (e.g., disinformation, generating fake profiles, surveillance), fairness considerations (e.g., deployment of technologies that could make decisions that unfairly impact specific groups), privacy considerations, and security considerations.
        \item The conference expects that many papers will be foundational research and not tied to particular applications, let alone deployments. However, if there is a direct path to any negative applications, the authors should point it out. For example, it is legitimate to point out that an improvement in the quality of generative models could be used to generate Deepfakes for disinformation. On the other hand, it is not needed to point out that a generic algorithm for optimizing neural networks could enable people to train models that generate Deepfakes faster.
        \item The authors should consider possible harms that could arise when the technology is being used as intended and functioning correctly, harms that could arise when the technology is being used as intended but gives incorrect results, and harms following from (intentional or unintentional) misuse of the technology.
        \item If there are negative societal impacts, the authors could also discuss possible mitigation strategies (e.g., gated release of models, providing defenses in addition to attacks, mechanisms for monitoring misuse, mechanisms to monitor how a system learns from feedback over time, improving the efficiency and accessibility of ML).
    \end{itemize}
    
\item {\bf Safeguards}
    \item[] Question: Does the paper describe safeguards that have been put in place for responsible release of data or models that have a high risk for misuse (e.g., pre-trained language models, image generators, or scraped datasets)?
    \item[] Answer: \answerNA{}
    \item[] Justification: The key released assets are a biomechanical simulation body, a neural circuit model, and simulation and training infrastructure; they pose no foreseeable misuse risk.
    \item[] Guidelines:
    \begin{itemize}
        \item The answer \answerNA{} means that the paper poses no such risks.
        \item Released models that have a high risk for misuse or dual-use should be released with necessary safeguards to allow for controlled use of the model, for example by requiring that users adhere to usage guidelines or restrictions to access the model or implementing safety filters. 
        \item Datasets that have been scraped from the Internet could pose safety risks. The authors should describe how they avoided releasing unsafe images.
        \item We recognize that providing effective safeguards is challenging, and many papers do not require this, but we encourage authors to take this into account and make a best faith effort.
    \end{itemize}

\item {\bf Licenses for existing assets}
    \item[] Question: Are the creators or original owners of assets (e.g., code, data, models), used in the paper, properly credited and are the license and terms of use explicitly mentioned and properly respected?
    \item[] Answer: \answerYes{}
    \item[] Justification: Third-party assets are cited; their licenses are permissive and compatible with our use.
    \item[] Guidelines:
    \begin{itemize}
        \item The answer \answerNA{} means that the paper does not use existing assets.
        \item The authors should cite the original paper that produced the code package or dataset.
        \item The authors should state which version of the asset is used and, if possible, include a URL.
        \item The name of the license (e.g., CC-BY 4.0) should be included for each asset.
        \item For scraped data from a particular source (e.g., website), the copyright and terms of service of that source should be provided.
        \item If assets are released, the license, copyright information, and terms of use in the package should be provided. For popular datasets, \url{paperswithcode.com/datasets} has curated licenses for some datasets. Their licensing guide can help determine the license of a dataset.
        \item For existing datasets that are re-packaged, both the original license and the license of the derived asset (if it has changed) should be provided.
        \item If this information is not available online, the authors are encouraged to reach out to the asset's creators.
    \end{itemize}

\item {\bf New assets}
    \item[] Question: Are new assets introduced in the paper well documented and is the documentation provided alongside the assets?
    \item[] Answer: \answerYes{}
    \item[] Justification: Rat~v2 and Rat~NCAP are documented and released with code at \website.
    \item[] Guidelines:
    \begin{itemize}
        \item The answer \answerNA{} means that the paper does not release new assets.
        \item Researchers should communicate the details of the dataset\slash code\slash model as part of their submissions via structured templates. This includes details about training, license, limitations, etc. 
        \item The paper should discuss whether and how consent was obtained from people whose asset is used.
        \item At submission time, remember to anonymize your assets (if applicable). You can either create an anonymized URL or include an anonymized zip file.
    \end{itemize}

\item {\bf Crowdsourcing and research with human subjects}
    \item[] Question: For crowdsourcing experiments and research with human subjects, does the paper include the full text of instructions given to participants and screenshots, if applicable, as well as details about compensation (if any)? 
    \item[] Answer: \answerNA{}
    \item[] Justification: The paper does not involve crowdsourcing or human subjects.
    \item[] Guidelines:
    \begin{itemize}
        \item The answer \answerNA{} means that the paper does not involve crowdsourcing nor research with human subjects.
        \item Including this information in the supplemental material is fine, but if the main contribution of the paper involves human subjects, then as much detail as possible should be included in the main paper. 
        \item According to the NeurIPS Code of Ethics, workers involved in data collection, curation, or other labor should be paid at least the minimum wage in the country of the data collector. 
    \end{itemize}

\item {\bf Institutional review board (IRB) approvals or equivalent for research with human subjects}
    \item[] Question: Does the paper describe potential risks incurred by study participants, whether such risks were disclosed to the subjects, and whether Institutional Review Board (IRB) approvals (or an equivalent approval/review based on the requirements of your country or institution) were obtained?
    \item[] Answer: \answerNA{}
    \item[] Justification: The paper does not involve human subjects.
    \item[] Guidelines:
    \begin{itemize}
        \item The answer \answerNA{} means that the paper does not involve crowdsourcing nor research with human subjects.
        \item Depending on the country in which research is conducted, IRB approval (or equivalent) may be required for any human subjects research. If you obtained IRB approval, you should clearly state this in the paper. 
        \item We recognize that the procedures for this may vary significantly between institutions and locations, and we expect authors to adhere to the NeurIPS Code of Ethics and the guidelines for their institution. 
        \item For initial submissions, do not include any information that would break anonymity (if applicable), such as the institution conducting the review.
    \end{itemize}

\item {\bf Declaration of LLM usage}
    \item[] Question: Does the paper describe the usage of LLMs if it is an important, original, or non-standard component of the core methods in this research? Note that if the LLM is used only for writing, editing, or formatting purposes and does \emph{not} impact the core methodology, scientific rigor, or originality of the research, declaration is not required.
    \item[] Answer: \answerNA{}
    \item[] Justification: LLMs were not used as part of the core methods.
    \item[] Guidelines:
    \begin{itemize}
        \item The answer \answerNA{} means that the core method development in this research does not involve LLMs as any important, original, or non-standard components.
        \item Please refer to our LLM policy in the NeurIPS handbook for what should or should not be described.
    \end{itemize}

\end{enumerate}

\end{document}